\documentclass[onecolumn,fleqn,usenatbib]{mnras}

\usepackage[british]{babel}

\usepackage{soul}

\usepackage{amsbsy}
\usepackage{amssymb}
\usepackage{amsmath}
\usepackage{bm}
\usepackage{xfrac}
\usepackage{physics}
\usepackage{multirow}

\usepackage{graphicx}

\hypersetup{
    bookmarksnumbered=true,
    breaklinks=false,
    citecolor=blue,
    colorlinks=true,
}

\usepackage[capitalise,compress]{cleveref}

\crefname{section}{Sec.}{Secs.}
\Crefname{section}{Section}{Sections}

\begin{document}

\title{Neutron star g-modes in the relativistic Cowling approximation}

\author[Counsell et al] {A. R. Counsell$^{1}$, F. Gittins$^{1}$,  N. Andersson$^{1}$ \& P. Pnigouras$^{2}$ \\
$^{1}$Mathematical Sciences and STAG Research Centre, University of Southampton, Southampton SO17 1BJ, United Kingdom\\
$^{2}$Departamento de F\'isica Aplicada, Universidad de Alicante, Campus de San Vicente del Raspeig, Alicante 03690, Spain
    }

\maketitle

\begin{abstract}
Mature neutron stars are expected to exhibit gravity g-modes due to stratification caused by a varying matter composition in the high-density  core. By employing the BSk equation of state family, and working within the relativistic Cowling approximation, we examine how subtle differences in the nuclear matter assumptions impact on the g-mode spectrum. We investigate the possibility of detecting individual g-mode resonances during a binary inspiral with current and next-generation ground-based detectors, like Cosmic Explorer and the Einstein Telescope. Our results suggest that these resonances may be within the reach of future detectors, especially for low mass stars with  $M\lesssim1.4M_\odot$.
\end{abstract}

\section{Introduction}

Neutron stars are highly compact and complex objects, the description of which requires  a rich variety of physics. In particular, there are specific classes of oscillation modes associated with each aspect of the physics involved. One such feature is the varying composition of matter throughout the neutron star, introducing buoyancy as a restoring force in the equations of fluid dynamics \citep{Reisenegger}. The buoyancy gives rise to  low-frequency ($\lesssim 1$ kHz) gravity g-modes. A realistic description of these modes requires both Einstein's general relativity and a detailed prescription for the strong nuclear interactions that occur at supernuclear densities. 

Given that the dense nuclear matter equation of state is largely unknown, it is in fact the inverse problem that we are interested in. We want to establish how observational data can be used to deduce/constrain the neutron star equation of state. The techniques of stellar seismology for neutron stars are particularly promising in this respect; a star's oscillation features provide a unique probe of the neutron star interior \citep{McDermott,McDermottVanHorn,UniversalRelations, Miniutti}. Interest in this problem was naturally elevated by the  detection of gravitational waves from the  GW170817 neutron star merger event \citep{GW170817,GW170817-2}, and the problem is under intense scrutiny given the proposals for a next generation of ground-based gravitational-wave instruments, Cosmic Explorer \citep{CosmicExplorer} and the Einstein Telescope \citep{ET}. These future instruments are expected to (finally) open a window to the neutron star oscillation spectrum. 

Much of the literature on neutron star seismology has focused on oscillation modes that depend weakly on the precise nuclear physics below the crust, such as the fundamental f-mode or the pressure p-modes. There has also been a considerable amount of work on the gravity g-modes (such as \citealt{McDermott}, \citealt{Finn} and \citealt{Miniutti}), which should be present in both cold mature and hot young neutron stars \citep[noting  evidence from numerical simulations that the g-modes may be excited during the proto-neutron star stage following a core-collapse supernova;][]{Burrows}. In general, the g-modes depend on both the internal matter composition and the state of matter, so it is important to understand what constraints on the nuclear physics could be made from  observations. 

Focusing on a mature neutron star, the g-mode frequencies depend on the variation of the lepton fraction with density (and the state of matter, like the presence of superfluidity; \citealp{2014MNRAS.442L..90K,2016MNRAS.455.1489P}), which is sensitive to nuclear parameters such as the nuclear symmetry energy (defined as the difference between the energy per nucleon in pure neutron matter and the energy per nucleon in symmetric nuclear matter). Therefore, an observation of a specific g-mode frequency, for example, manifesting as a tidal resonance during binary inspiral \citep{DongLai,Kokkotas95,Wynn,Wynn2}, could provide information about the underlying nuclear physics and the equation of state beyond bulk properties such as the mass and radius of the star.

In this paper we examine
how uncertainties in the underlying nuclear physics impact on the matter  composition and, in turn, affect the g-mode frequencies and their detectability. Specifically, we consider the spectrum of g-modes in a cold mature neutron star based on different equations of state from the BSk family \citep{BSk,BSk2,BSk3}, specifically: BSk22, BSk24, BSk25 and BSk26 \citep{BSkGR,BSkHFB}. These equations of state are based on generalised Skyrme-type forces with parameters fitted to nuclear and astrophysical data. 

The advantage of the BSk family of models is that it allows us to employ a realistic description of the matter stratification, going beyond previous work that tends to assume a constant adiabatic index for the perturbations ($\Gamma_1$ later). While the same information can be extracted for  other “realistic” equation of state models, the BSk family has the advantage of being expressed in closed form, which means that the thermodynamical derivatives required for the g-mode calculation do not have to be worked out numerically (as done in, for example,  the recent work by \citet{gittins2024}). An analytic representation is convenient as it avoids  numerical errors that would be unavoidable if we were to base the calculation on tabulated equation of state data.

The layout of the paper is as follows: in Section \ref{Method}, we outline the background equations and physics that go into the problem. The Cowling approximation, in which spacetime perturbations are neglected, is used and the equations are shown to be in the same form as in \citet{McDermott}. Section \ref{Results} presents and discusses our numerical results along with estimates of the detectability of tidal resonances. Finally, Section \ref{Conclusions} summarises the work and presents ideas for future continuation of this effort.

Throughout the paper, spatial indices are denoted with Latin characters $i, j, k, \ldots$ and spacetime indices are denoted with early Latin characters $a, b, c, \ldots$. The indices $l$ and $m$ will be used exclusively for spherical-harmonic multipoles. The Einstein summation convention will be used for repeated tensor component indices.  The signature of the spacetime metric $g_{ab}$ is $(-,+,+,+)$. 

\section{The Perturbation Problem }\label{Method}

We want to study the oscillation properties of a non-rotating relativistic star. This problem has been explored in great detail in the literature and the various steps required in its formulation are well known. Nevertheless, we will provide the key ingredients here, partly for completeness but also because the discussion provides a clear illustration of the various simplifying assumptions we introduce.

The first step involves establishing the background model. For reasons which will become clear later, the next few steps will be taken without committing to a specific choice of spatial coordinates. Later, the conventional choice of Schwarzschild coordinates will be introduced. 

\subsection{The fluid equations}

Initially, progress can be made by assuming that the unperturbed metric is obtained from 
\begin{equation}
    ds^2 = - e^\nu c^2 dt^2 +  \gamma_{ij} dx^i dx^j  \ .
\end{equation}
where the spatial part of the metric, $\gamma_{ij}$, is diagonal. Assuming a perfect fluid, the stress-energy tensor takes the form
\begin{equation}
    T^{ab} = {1\over c^2} \left( \varepsilon+ p\right) u^a u^b + p g^{ab} = {\varepsilon \over c^2} u^a u^b + p\perp^{ab}, 
\end{equation}
where $\varepsilon$ is the energy density, $p$  represents the pressure, the fluid four velocity $u^a$ is normalised in such a way that 
\begin{equation}
    u_a u^a = g_{ab} u^a u^b = - c^2,
\end{equation}
and the orthogonal projection is defined by 
\begin{equation}
    \perp^{ab} = g^{ab} + {1\over c^2} u^a u^b,
\end{equation}
with $g^{ab}$  the  inverse metric.

In order to write down the equations of fluid dynamics one needs
\begin{equation}
  \nabla_b T^{ab} 
    = {1\over c^2} \left[ u^a \nabla_b \left( \varepsilon u^b \right) +  \varepsilon u^b \nabla_b u^a \right] + \perp^{ab} \nabla_b p +  {p \over c^2} \left( u^a \nabla_b u^b + u^b \nabla_b u^a  \right) = 0.
    \label{divT}
\end{equation}
Projecting along $u_a$, one obtains the energy equation:
\begin{equation}
    u^b \partial_b \varepsilon + (p+\varepsilon) \nabla_b u^b = 0.
\end{equation}
Meanwhile, the orthogonal projection of \eqref{divT} leads to the momentum equation:
\begin{equation}
    {p+\varepsilon \over c^2} u^b \nabla_b u^c + \perp^{cb} \partial_b p = 0 .
\end{equation}
Lastly, we need the equation for baryon number conservation,
\begin{equation}
\nabla_a \left( n u^a\right) = 0,
\label{BaryonCon}
\end{equation}
where $n$ is the baryon number density.
The system of equations is closed by an equation of state for matter, providing the pressure $p$ as a function of (say) the number density $n$, the matter composition and temperature. In our analysis, we focus on mature neutron stars, which are expected to be cold enough that thermal effect can be ignored. The equation of state can then be taken to be a function of two parameters: $n$ and the lepton (in our case, as we ignore the presence of muons etc, equal to the proton/electron) fraction. 

\subsection{Perturbations}

Next, we want to consider perturbations. For the velocity, we introduce the Eulerian perturbation (indicated by $\delta$) such that the total four-velocity is 
\begin{equation}
\bar u^a = u^a + \delta u^a
\Longrightarrow \bar u_a \bar u^a = \bar g_{ab} \left( u^a + \delta u^a \right) ( u^b + \delta u^b ) = - c^2 .
\end{equation}
The metric is similarly given by
\begin{equation}
   \bar g_{ab} =  g_{ab} + \delta g_{ab}. 
\end{equation}
This means that, to linear order, we have 
\begin{equation}
\delta u^0 = {c \over 2} e^{-3\nu/2}  \delta g_{00}. 
\end{equation}
Evidently, the normalisation does not involve the spatial components $\delta u^i$. Rather, it fixes the time component so that the perturbed four-velocity has three components, just like in classical fluid dynamics. 

Next, we introduce metric perturbations ``inspired'' by the ADM formalism \citep{ADM}. That is, 
\begin{gather} 
\delta g_{00} = \delta \alpha,
\\
\delta g_{0i} =\delta g_{i0} = \delta\beta_i,
\\
\delta g_{ij} = \delta \gamma_{ij}.
\end{gather}
These variables will be further expanded in angular harmonics later.
This means that we have
\begin{equation}
  \delta u^0 = {c \over 2} e^{-3\nu/2}  \delta \alpha.
\end{equation}

Now let us return to the perturbation equations. Linearising \eqref{BaryonCon}, it  follows that 
\begin{equation}
 \partial_t \delta n  +  {e^{\nu/2} \over \sqrt{-g}}\partial_i \left[ \sqrt{-g} n \delta u^i \right] = -  {n\over 2} e^{-\nu} \partial_t \delta \alpha,
\end{equation}
where $g=$ det$(g_{ab})$.
Moving on to the perturbed energy equation, we have
\begin{equation}
  \left( \sqrt{-g} u^b\right) \partial_b \varepsilon + (p+\varepsilon) \partial_b \left( \sqrt{-g} u^b \right)  = 0.
\end{equation}
It follows that 
\begin{equation}
   \sqrt{-g} e^{-\nu/2} \left[ \partial_t \delta \varepsilon + {1\over 2}(p+\varepsilon) e^{-\nu} \partial_t \delta \alpha \right] +  \sqrt{-g} \delta u^j \partial_j  \varepsilon + (p+\varepsilon) \partial_i \left( \sqrt{-g} \delta u^i \right) = 0,
\end{equation}
which leads to
\begin{equation}
  \partial_t \delta \varepsilon + {1\over 2}(p+\varepsilon) e^{-\nu} \partial_t \delta \alpha + e^{\nu/2}\delta u^i \partial_i  \varepsilon  + (p+\varepsilon){1\over \sqrt{\gamma}} \partial_i \left[e^{\nu/2} \sqrt{\gamma} \delta u^i \right] = 0,
  \label{GREnergy}
\end{equation}
where $\gamma = $ det$(\gamma_{ij})$.
Finally, the perturbed momentum equation is given by
\begin{equation}
{p+\varepsilon \over c^2} e^{-\nu/2}   \partial_t \delta u^i +  {1\over 2} (\delta p+\delta \varepsilon) g^{ij} \partial_j \nu 
+  g^{ij} \partial_j \delta p -  {p+\varepsilon \over 2}  e^{-\nu} (g^{ij} \partial_j \nu) \,\delta \alpha - {p+\varepsilon\over 2}     g^{ij}e^{-\nu}  \left( {2 \over c}  \partial_t \delta \beta_j  - \partial_j \delta \alpha \right) = 0.
 \label{GRMomentum}
\end{equation}

At this point, it is notable that there is no direct coupling to $\delta g_{ij}$ in the final momentum equation. Any coupling to the spatial part of the metric enters via the Einstein equations. This observation is useful for several reasons. For example, it helps explain how different versions of the relativistic Cowling approximation---e.g. whether we assume that all components of the perturbed metric are ignored (as in \citealt{McDermott}) or retain the ``momentum parts'' (as advocated by \citealt{FinnCowling})---impact on the problem. Moreover, the absence of $\delta g_{ij}$ in \eqref{GRMomentum} may help explain why oscillation modes obtained in the conformal flatness approximation can be quite accurate, as demonstrated by, for example, \citet{2019MNRAS.482.3967T}.  These issues would be worth closer inspection, but we will not explore them further here.

\subsection{Cowling Approximation}

Having derived the perturbation equations of a relativistic, non-rotating star, we now want to simplify the problem by introducing the (relativistic) Cowling approximation. Here we take this to mean that we ignore the perturbations of the gravitational field \citep{McDermott}. For relativistic problems, this obviously involves omitting the gravitational-wave aspects. This is not expected to be a good approximation for oscillations that are efficient gravitational-wave emitters (like the fundamental f-mode or, indeed, the spacetime w-modes \citep{1992MNRAS.255..119K,1996MNRAS.280.1230A}, the latter of which do not even exist in the Cowling approximation). However, the gravity g-modes are weakly damped by gravitational-wave emission, so the approximation should be adequate for the corresponding low-frequency dynamics \citep{FinnCowling,ChristianThesis}.

We introduce the Cowling approximation  by setting all metric perturbations to zero, i.e.
\begin{equation}
    \delta \alpha = \delta \beta_i = 0.
\end{equation}
The accuracy of this approximation is discussed in detail in \citet{FinnCowling}, with particular focus on the g-modes.
Along with this, we introduce the fluid displacement vector, $\xi^i$, defined as 
\begin{equation}
    \delta u^i = ce^{-\nu/2}\partial_0 \xi^i=e^{-\nu/2}\partial_t \xi^i,
\end{equation}
which follows from our assumed gauge condition. With this definition, equation~\eqref{GRMomentum} leads to
\begin{equation}
    {p+\varepsilon \over c^2} e^{-\nu}   \partial^2_t  \xi_i +  {1\over 2} (\delta p+\delta \varepsilon)  \partial_i \nu 
+  \partial_i \delta p = 0.
\label{CowlingMomentum}
\end{equation}

Now, considering the problem for non-rotating stars, we assume that the oscillation modes, with label $n$ and frequency $\omega_n$, are associated with a polar perturbation displacement vector. In terms of the coordinate basis associated with the spherical polar coordinates $[r,\theta,\varphi]$, we then have (noting that the mode problem is degenerate in the azimuthal angle $\varphi$ for spherical stars so it is sufficient to keep track of the polar angle harmonic index $l$)
\begin{equation}
    \xi^i(t,r,\theta,\varphi) =  \xi_l^i (r,\theta,\varphi) e^{i\omega_n t} \ ,
\end{equation}
with spatial part 
\begin{equation}
  \xi_l^i =  { 1 \over r} W_l Y_l^m {\delta}^i_r + { 1 \over r^2}  V_l \partial_\theta Y_l^m {\delta}_\theta^i + { 1 \over r^2 \sin^2 \theta}    V_l Y_l^m  {\delta}_\varphi^i  \ .
  \label{static xi}
\end{equation}
The radial and angular amplitudes, $W_l$ and $V_l$, are functions of $r$ only. Along with this, all scalar perturbations are expanded in spherical harmonics. For example,  the perturbed pressure is
\begin{equation}
    \delta p = \delta  p_l Y_l^m  e^{i\omega_n t},
\end{equation}
and similar for $\delta \varepsilon$. Finally, the background metric is taken to have the Schwarzschild form
\begin{equation}
    ds^2 = - e^\nu c^2 dt^2 + e^\lambda dr^2 +r^2 d\theta^2 + r^2\sin^2\theta d\varphi^2  \ ,
\end{equation}
which means that the background configuration is obtained by solving the standard Tolman-Oppenheimer-Volkoff equations.

Substituting the different expressions into \eqref{CowlingMomentum} gives the following: For the $r$-component we get
\begin{equation}
    \frac{p+\varepsilon}{c^2}\omega_n^2e^{\lambda-\nu} \frac{W_l}{r}= \frac{\delta p_l + \delta \varepsilon_l}{2} {d\nu \over dr} + \partial_r \delta p_l ,
\end{equation}
while  the $\varphi$-component leads to
\begin{equation}
\frac{p+\varepsilon}{c^2}\omega_n^2e^{-\nu} V_l= \delta p_l.
\end{equation}

Next, applying the Cowling approximation to the perturbed energy equation \eqref{GREnergy} gives
\begin{equation}
  \partial_t \delta \varepsilon + e^{\nu/2}\delta u^i \partial_i  \varepsilon  + (p+\varepsilon)  {1\over \sqrt{\gamma}} \partial_i \left[e^{\nu/2} \sqrt{\gamma} \delta u^i \right] = 0 \ , 
  \label{delepsilon}
\end{equation}
where
\begin{equation}
    \sqrt{\gamma} = e^{\lambda/2} r^2 \sin\theta.
\end{equation}
In terms of the defined variables, this becomes
\begin{equation}
  \partial_t \delta \varepsilon + e^{\nu/2}\delta u^r \partial_r  \varepsilon  + (p+\varepsilon)  {1\over \sqrt{\gamma}} \partial_r \left[e^{\nu/2} \sqrt{\gamma} \delta u^r \right]+ (p+\varepsilon)  {1\over \sqrt{\gamma}} \partial_\theta \left[e^{\nu/2} \sqrt{\gamma} \delta u^\theta \right]  + (p+\varepsilon)  {1\over \sqrt{\gamma}} \partial_\varphi \left[e^{\nu/2} \sqrt{\gamma} \delta u^\varphi \right]= 0 \ , 
\end{equation}
which can be simplified to
\begin{equation}
{ p+\varepsilon  \over r^2} \left[  \partial_r \left( r W_l\right)  - l(l+1)  V_l \right] +  \left( {d \varepsilon \over dr}   + { p+\varepsilon  \over 2} {d\lambda \over dr}\right) {W_l\over r}  
   + \delta \varepsilon_l   = 0 \ .
\end{equation}

Putting all these equations together, we have
\begin{equation}
    \frac{p+\varepsilon}{c^2}\omega_n^2e^{\lambda-\nu} \frac{W_l}{r}= \frac{\delta p_l + \delta \varepsilon_l}{2} {d\nu \over dr} + \partial_r \delta p_l ,
\end{equation}
\begin{equation}
\frac{p+\varepsilon}{c^2}\omega_n^2e^{-\nu} V_l= \delta p_l,
\label{deltap}
\end{equation}
and
\begin{equation}
{ p+\varepsilon  \over r^2} \left[  \partial_r \left( r W_l\right)  - l(l+1)  V_l \right] +  \left( {d \varepsilon \over dr}   + { p+\varepsilon  \over 2} {d\lambda \over dr}\right) {W_l\over r}  
   + \delta \varepsilon_l   = 0 \ .
\end{equation}
In order to solve these equations we need to relate $\delta p$ and $\delta \varepsilon$. This requires information from the thermodynamics of the nuclear matter. 

Previously, the majority of the literature on neutron star oscillations assumes that, as the matter is cold, the fluid is barotropic and thus only involves one variable, i.e. $\varepsilon = \varepsilon(n)$. One can then use $p=p(n)$ to arrive at $p=p(\varepsilon)$, which is needed to close the perturbation equations. However, while this is a valid approximation for  cold neutron stars in equilibrium, it cannot be applied to g-modes. As first argued by \citet{Reisenegger}, the characteristic timescale associated with the relevant weak interactions is much longer than the typical oscillation periods of these modes. Therefore, as a fluid element in the star is displaced from its equilibrium position, nuclear reactions do not act fast enough to restore $\beta$ equilibrium between the element and its new environment. Thus, buoyancy forces associated with the differing composition cause the displaced element to oscillate, giving rise to the composition g-modes that we are interested in.
In general, we  assume that the composition is frozen during the oscillation, i.e. $\Delta x_e = 0$ where $x_e$ is the electron fraction and $\Delta$ is the Lagrangian perturbation (for a recent study on the g-mode spectrum in the
intermediate regime of finite reaction times, see \citealt{Rhys}). When thermal effect are accounted for, we  also assume no heat is transferred, i.e. $\Delta s = 0$, where $s$ the entropy per baryon. 

To close the system of equations we introduce the adiabatic index $\Gamma_1$ of the perturbed matter defined such that
\begin{equation}
    \Delta p = \frac{\Gamma_1 p}{\varepsilon+p}\Delta \varepsilon ,
\end{equation}
with
\begin{equation}
    \Gamma_1=\frac{p+\varepsilon}{p}\left(\frac{\partial p}{\partial \varepsilon}\right)_{s,x_e}
\end{equation}
holds information about composition and entropy gradients. In the models we consider here, the stars are taken to be cold enough that thermal aspects can be ignored.

Next, using the definition of the Lagrangian perturbation, we get
\begin{equation}
    \delta p = \frac{\Gamma_1 p}{\varepsilon+p} \left(\delta \varepsilon +\frac{W_l}{r} \frac{d\varepsilon}{dr} \right) - \frac{W_l}{r} \frac{dp}{dr}.
\end{equation}
From this relation we arrive at the differential equations for the amplitudes $W_l$ and $V_l$:
\begin{equation}
    \frac{dW_l}{dr} = \left[\frac{l(l+1)}{r}-\frac{\omega_n^2}{c^2}\frac{p+\varepsilon}{\Gamma_1 p}re^{-\nu}\right]V_l-\left(\frac{1}{r}+\frac{1}{2}\frac{d\lambda}{dr}+\frac{1}{\Gamma_1 p}\frac{dp}{dr}\right)W_l,
    \label{dwdr}
\end{equation}
and
\begin{equation}
    \frac{dV_l}{dr}=\left(e^{\lambda}-\frac{1}{p+\varepsilon}\frac{dp}{dr}\frac{c^2}{\omega_n^2}A_+e^{\nu+\lambda/2}\right)\frac{W_l}{r}-A_-e^{\lambda/2}V_l,
    \label{dvdr}
\end{equation}
where,
\begin{equation}
    A_\pm = e^{-\lambda/2}\left[\frac{1}{p+\varepsilon}\frac{d(p+\varepsilon)}{dr}-\frac{1}{\Gamma_1 p}\frac{dp}{dr}\left(1\pm\frac{\Gamma_1 p}{p+\varepsilon}\right)\right].
\end{equation}

Finally, in order to write the equations in a form better suited for numerical integration, we introduce new variables:
\begin{equation}
    Z_1 = \frac{W_l}{r^2}\left(\frac{r}{R}\right)^{2-l}e^{\lambda/2},
\end{equation}
and
\begin{equation}
    Z_2=\frac{\omega_n^2r}{GM_r}V_l\left(\frac{r}{R}\right)^{2-l},
\end{equation}
where $R$ is the stellar radius and $M_r$ is given by
\begin{equation}
    \frac{dM_r}{dr}=\frac{4\pi r^2\varepsilon}{c^2}.
\end{equation}
We  can then rewrite the differential equations as
\begin{equation}
    r\frac{d Z_1}{d r}=\left(\frac{V}{\Gamma_1}-l-1\right)Z_1+\left[l(l+1)\frac{GM_r}{\omega_n^2 r^3}e^{\lambda/2}-\frac{V}{\Gamma_1 \beta}\right]Z_2,
    \label{Z1}
\end{equation}
and
\begin{equation}
    r\frac{d Z_2}{d r}=\left(\frac{\omega_n^2r^3}{GM_r}+A_+r\beta\right)e^{\lambda/2}Z_1+\left(3-l-U-A_-e^{\lambda/2}r\right)Z_2,
    \label{Z2}
\end{equation}
where
\begin{equation}
    U=\frac{d \ln M_r}{d \ln r},
\end{equation}
\begin{equation}
    V=-\frac{d \ln p}{d \ln r},
\end{equation}
and
\begin{equation}
    \beta=e^{\nu+\lambda/2}\left(1+\frac{4\pi r^3p}{M_rc^2}\right).
\end{equation}
The equations are now identical to those used by \cite{McDermott}.

Two boundary conditions are  required for this system, one at the centre and one at the surface of the star. At the stellar surface one requires that $\Delta p = 0$. Using \eqref{deltap} this corresponds to
\begin{equation}
    Z_2-\beta Z_1=0.
\end{equation}
At the centre of the star, the differential equations \eqref{dwdr}--\eqref{dvdr} are required to be regular as $r \rightarrow 0$. This leads to
\begin{equation}
    Z_1-\frac{lGM_r}{r^3 \omega_n^2}Z_2=0.
\end{equation}
Lastly a normalisation condition for our solutions is added at the stellar surface. Specifically, we use
\begin{equation}
    Z_1e^{-\lambda/2}=1,
\end{equation}
which ensures that
\begin{equation}
    \frac{\xi^r(R)}{R}=1.
    \label{XiNorm}
\end{equation}

\subsection{Mode orthogonality}

Investigations into the g-mode properties are topical because of the possibility that the associated tidal resonances  may be detectable during the late stage of binary inspiral with future gravitational-wave detectors. In Newtonian gravity, the tidal response is represented by a mode sum \citep{1994MNRAS.270..611L,1995MNRAS.275..301K,2024MNRAS.527.8409P}, motivated by the fact that the individual mode solutions are orthogonal with respect to a specific inner product \citep{Friedman}.
While a corresponding mode-sum expression is not yet established in general relativity (and one may argue that it should not exist, at least not as an exact expression; \citealp{Poisson24}), the Cowling approximation allows us to make progress in this direction.  
In essence, we want to show that the mode equations are Hermitian for some inner product. If this is the case, one has a suitable basis for a mode expansion, which will prove useful later on. The argument is not exactly new, but it is useful to spell out the required steps given the fact that different versions of the result exist in the literature and we want to make sure that our model is internally consistent.   

First, we express the momentum equation \eqref{CowlingMomentum} as 
\begin{equation}
   -\omega_n^2  {p+\varepsilon \over c^2} e^{-\nu}    \xi_i +  {1\over 2} (\delta p+\delta \varepsilon)  \partial_i \nu 
+  \partial_i \delta p = 0,
\end{equation}
which can then be written as \citep{Friedman}
\begin{equation}
    -\omega_n^2 A\xi_i + C_{ij} \xi^j = 0 \ . 
\end{equation}
In order for this equation to be Hermitian, we need a suitable inner product
\begin{equation}
    \langle \eta^i ,  \xi_i \rangle = \int \eta^{i*} f \xi_i dV = \int \eta^{i*} f \xi_i \sqrt{-g} d^3x,
\end{equation}
where $\xi^i$ and $\eta^i$ are solutions to the perturbation equations and  $*$ denotes the complex conjugate.
Specifically, we need to identify a function $f(r)$, such that 
\begin{equation}
     \langle \eta^i ,  C_{ij} \xi^j \rangle =  \langle \xi^i , C_{ij} \eta^j \rangle^*,
     \label{Crelation}
\end{equation}
and similarly for $A$. Once we establish this result, we can  define the symplectic product
\begin{equation}
    W(\eta^i,\xi_i) = \langle \eta^i, A \partial_t \xi_i\rangle - \langle A\partial_t\eta^i, \xi_i\rangle,
\end{equation}
such that
\begin{equation}
\partial_t W = \langle \eta^i, A\partial_t^2 \xi_i \rangle - \langle A\partial_t^2 \eta^i , \xi_i\rangle = -\langle \eta^i, C\xi_i\rangle + \langle C\eta^i , \xi_i\rangle  = - \langle \eta^i, C\xi_i\rangle + \langle \xi^i, C\eta_i\rangle^* = 0.  
\end{equation}
This demonstrates that $W$ provides a conserved quantity.
Moreover, for two mode solutions, $\xi^i e^{i\omega_n t}$ and $\eta^i e^{i\omega_{n'} t}$, we have
\begin{equation}
    \left( \omega_{n'}^2 - \omega_n^2\right) \left( \langle \eta^i, A\xi_ i\rangle + \langle A\eta^i, \xi_i\rangle\right) e^{i(\omega_n-\omega_{n'})t } = 0.
\end{equation}
Since $A$ and $f(r)$ are both real, it is easy to see that one must have
\begin{equation}
    \left( \omega_{n'}^2 - \omega_n^2\right)  \langle \eta^i_, A\xi_i\rangle e^{i(\omega_n-\omega_{n'})t } = 0.
\end{equation}
Assuming that the modes are not degenerate, this means that
\begin{equation}
\langle \eta^i, A\xi_i\rangle = \mathcal A^2_n \delta_{nn'},
\label{NormCondition}
\end{equation}
for some amplitude $\mathcal A^2_n$. This would then provide a basis for a mode expansion (following the steps from the Newtonian analysis).

In order to make progress towards \eqref{Crelation}, the starting point is
\begin{equation}
    \langle \eta^i, C_{ij} \xi^j \rangle = \int \eta^{i *} \left[ \frac{1}{2} (\delta_\xi \varepsilon + \delta_\xi p) \partial_i \nu + \partial_i \delta_\xi p \right] f \sqrt{- g} \, d^3 x,
\end{equation}
where $\delta_\xi$ indicates the perturbation associated with mode $\xi^i$.
Integrate by parts to find
\begin{equation}
    \langle \eta^i, C_{ij} \xi^i \rangle = \oint \eta^{i *} \delta_\xi p f \sqrt{- g} \, dS_i + \int \left[ \frac{1}{2} (\delta_\xi \varepsilon + \delta_\xi p) \eta^{i *} \partial_i \nu f \sqrt{- g} - \delta_\xi p \partial_i (\eta^{i *} f \sqrt{- g}) \right] \, d^3 x,
\end{equation}
where $dS_i$ is an outward-facing vectorial two-surface element arrived at through the divergence theorem. At the surface, $\delta_\xi p$ vanishes, hence we have
\begin{equation}
    \langle \eta^i, C_{ij} \xi^j \rangle = \int \left[ \frac{1}{2} (\delta_\xi \varepsilon + \delta_\xi p) \eta^{i *} \partial_i \nu f \sqrt{- g} - \delta_\xi p \partial_i (\eta^{i *} f \sqrt{- g}) \right] \, d^3 x.
\end{equation}
To keep this as general as possible, we require
\begin{equation}
    \frac{\Delta \varepsilon}{\varepsilon + p} = \frac{\Delta n}{n} = - \frac{1}{\sqrt{- g}} \partial_i (\sqrt{- g} \xi^i) - \frac{1}{2} \xi^i \partial_i \nu.
\end{equation}

Therefore,
\begin{gather}
   \frac{\delta \varepsilon}{\varepsilon + p} = - \frac{1}{\sqrt{- g}} \partial_i (\sqrt{- g} \xi^i) - \frac{1}{\varepsilon + p} \left( 1 + \frac{\varepsilon + p}{\Gamma p} \right) \xi^i \partial_i p, \\
   \frac{\delta p}{\Gamma_1 p} = - \frac{1}{\sqrt{- g}} \partial_i (\sqrt{- g} \xi^j) - \frac{1}{\Gamma_1 p} \left( 1 + \frac{\Gamma_1 p}{\varepsilon + p} \right) \xi^i \partial_i p,
\end{gather}
where
\begin{equation}
   \Gamma = \frac{\varepsilon + p}{p} \frac{dp}{d\varepsilon},
\end{equation}
represents the adiabatic index of the background configuration.

Thus, the inner product becomes
\begin{equation}
\begin{split}
    \langle \eta^i, C_{ij} \xi^j \rangle = &\int \frac{1}{\sqrt{- g}} \left( 1 + \frac{\Gamma_1 p}{\varepsilon + p} \right) \left[ \partial_j (\sqrt{- g} \xi^j) \eta^{j *} \partial_i p + \xi^j \partial_j p \partial_i (\sqrt{- g} \eta^{i *}) \right] f \sqrt{- g} \, d^3 x \\
    &+ \int \frac{1}{\varepsilon + p} \left( 2 + \frac{\varepsilon + p}{\Gamma p} + \frac{\Gamma_1 p}{\varepsilon + p} \right) \xi^j \partial_j p \eta^{i *} \partial_i p f \sqrt{- g} \, d^3 x + \int \frac{\Gamma_1 p}{(\sqrt{- g})^2} \partial_j (\sqrt{- g} \xi^j) \partial_i (\sqrt{- g} \eta^{i *}) f \sqrt{- g} \, d^3 x \\
    &+ \int \Gamma_1 p \left[ \frac{1}{\sqrt{- g}} \partial_j (\sqrt{- g} \xi^j) + \frac{1}{\Gamma_1 p} \left( 1 + \frac{\Gamma_1 p}{\varepsilon + p} \right) \xi^j \partial_j p \right] \eta^{i *} \frac{\partial_i f}{f} f \sqrt{- g} \, d^3 x.
\end{split}
\end{equation}
Evidently, this is Hermitian only when $f = \text{const}$. In view of this, we take $f = 1$ in the following.

Going back to \eqref{NormCondition}, we now have
 \begin{equation}
\langle \eta^i, A \xi_i \rangle  = \int \frac{p+\varepsilon}{c^2}  e^{-\nu} \eta^{i*} \xi_i \sqrt{-g}\, d^3x = \mathcal A^2_n \delta_{nn'},
\label{FullModeNorm}
\end{equation}
which shows  how the mode solutions $\xi^i$ should be normalised. For a single mode, substituting in \eqref{static xi}, this simplifies to
\begin{equation}
    \mathcal{A}^2_n=\int_{0}^{R}e^{(\lambda-\nu)/2}\frac{(\varepsilon+p)}{c^2}\left[W_l^2+\frac{l(l+1)}{r^2}V_l^2\right]d r,
    \label{A2Norm}
\end{equation}
in agreement with the result from \citet{Ipser} and the expression employed by \citet{Kokkotas}.

\section{Results}\label{Results}

The use of the Cowling approximation will (obviously) affect the accuracy of the mode solutions. As we are ignoring the dynamical aspects of spacetime---the gravitational-wave degrees of freedom---and also assume an ideal fluid---ignoring viscosity---the mode frequencies will be real valued. In full general relativity, the mode frequencies are complex as a result of damping due the emission of gravitational radiation (the modes satisfy a pure outgoing-wave condition at spatial infinity, see \citealt{1995MNRAS.274.1039A}). However, as we focus our attention on the g-modes, the frequencies of which have been shown to have very small imaginary parts, $\mathrm {Re}\, \omega_n \gg \mathrm{Im}\,\omega_n\approx0$ \citep{ChristianThesis}, the Cowling approximation is not expected to have a significant impact on the results. 

\subsection{Stellar models and the f-modes}\label{f-mode}

\begin{figure}
    \centering
       \includegraphics[height=6cm]{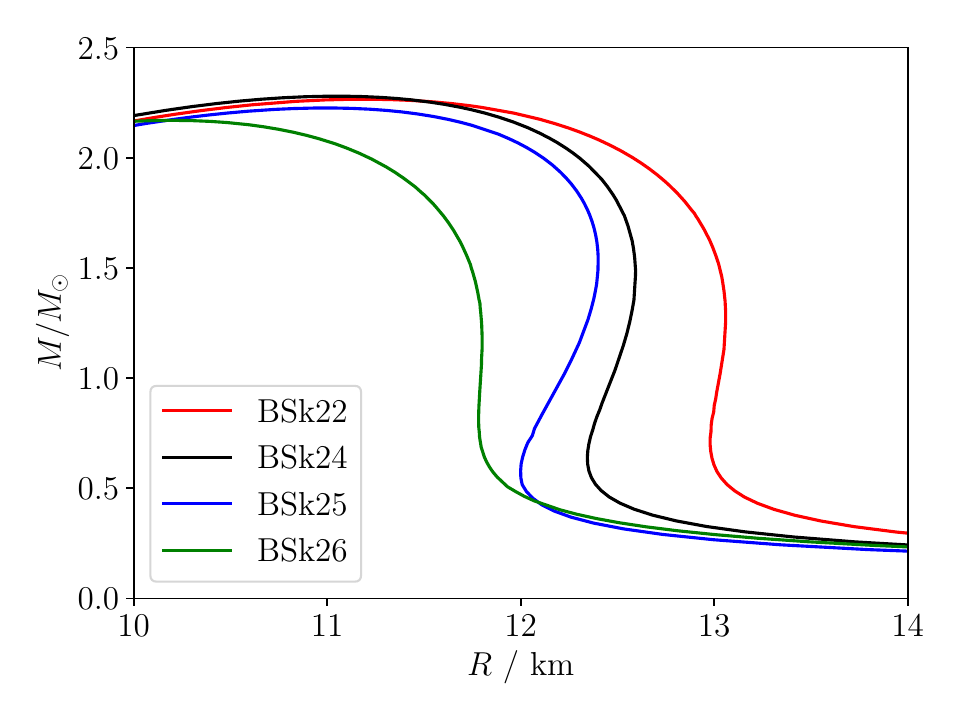}
    \includegraphics[height=6cm]{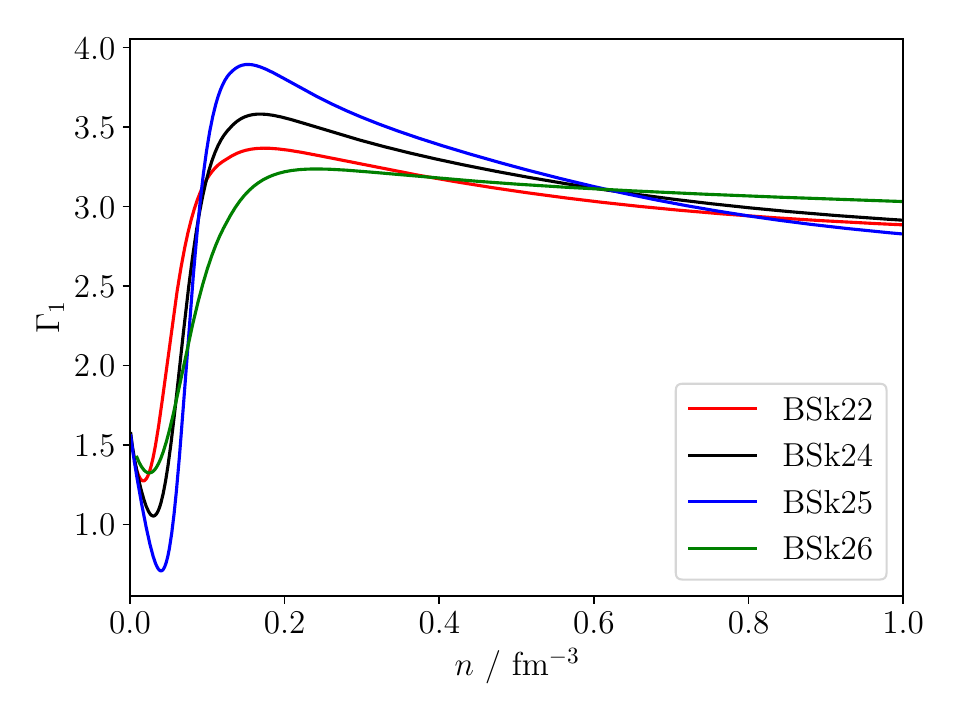}
    \caption{Left: Mass-radius curves for the chosen models from the BSk family (BSk22-26).
    Right: Plot showing the adiabatic index for frozen composition perturbations, $\Gamma_1$, vs baryon number density $n$.}
    \label{BSkGam1s}
\end{figure}

As already mentioned, we have opted to base our analysis on the BSk equation-of-state family \citep{BSkGR,BSkHFB}. Specifically, four different models will be compared against one another (BSk22; BSk24; BSk25 and BSk26). Mass-radius curves for these models are provided in the left panel of Figure~\ref{BSkGam1s}. Each of these equations of state are obtained from an analytical energy functional (see \citealt{BSkGR}), which can be used to calculate all necessary thermodynamic derivatives---required to, for example, work out the adiabatic index for frozen composition perturbations, $\Gamma_1$---and each model involves slightly different assumptions regarding the nuclear interactions. The adiabatic index is calculated following the steps laid out by \citet{FabianRmodes2}. Strictly speaking, this calculation only applies to the neutron star core. The relevant results are  provided in the right panel of Figure~\ref{BSkGam1s}, which shows $\Gamma_1$ for each equation of state as function of the baryon number density $n$,  clearly demonstrating that the common assumption of a constant $\Gamma_1$ is not appropriate for neutron stars. 

\begin{figure}
    \centering
    \includegraphics[height=6cm]{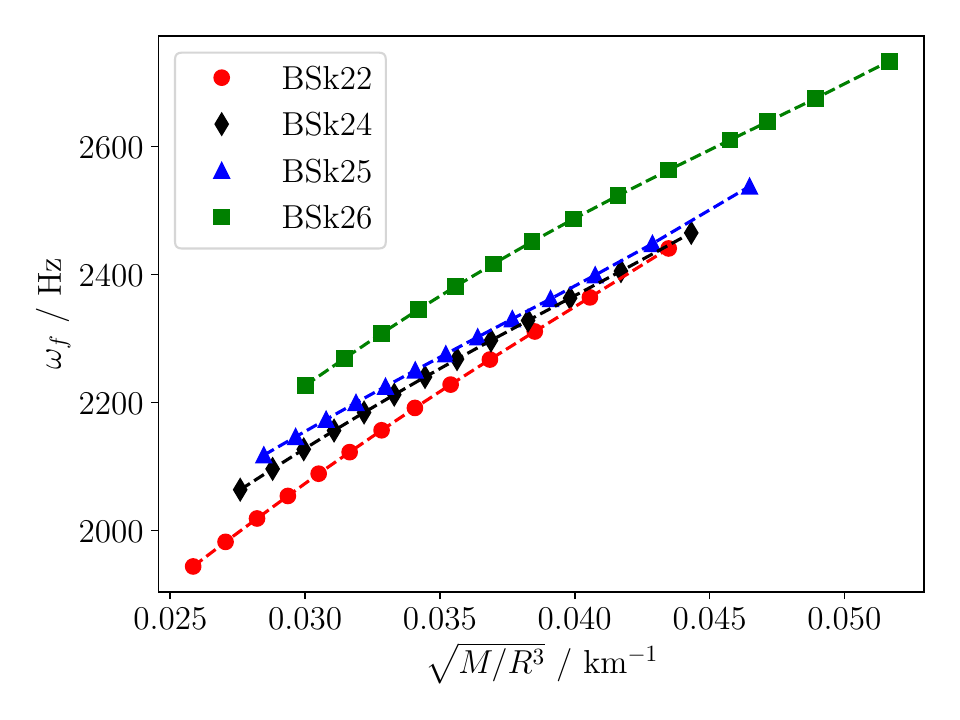}
    \caption{Plot of the f-mode frequency $\omega_f$ (in geometric units) vs $\sqrt{M/R^3}$ for BSk22-26 for total mass energies ranging from $M=1M_\odot$ up to the maximum allowed mass in each case.}
    \label{fmodeuniversal}
\end{figure}

Using each of the four equation of state models, equations \eqref{Z1} and \eqref{Z2} were solved for a variety of neutron star models with gravitational masses ranging from $M=1M_\odot$ up to the maximum allowed mass given by \citet{BSkGR}. 
In order to test the implementation, we first of all calculated the f-mode frequency, $\omega_f$.  The results (all  mode results in this paper are for $l=2$) are plotted, in geometric units  ($G = c = 1$), against $\sqrt{M/R^3}$ in Figure \ref{fmodeuniversal}. It is well known  \citep{UniversalRelations}, that in neutron stars 
$\omega_f \propto \sqrt{M/R^3}$ and our results bring out this expectation. 
However, we need to keep in mind that,  due to the Cowling approximation these frequencies will only be accurate to within about 15\% or so \citep{CowlingAccuracy}, hence the results should be considered with this in mind.

\subsection{The gravity g-modes}

\begin{figure}
    \centering
    \includegraphics[height=6cm]{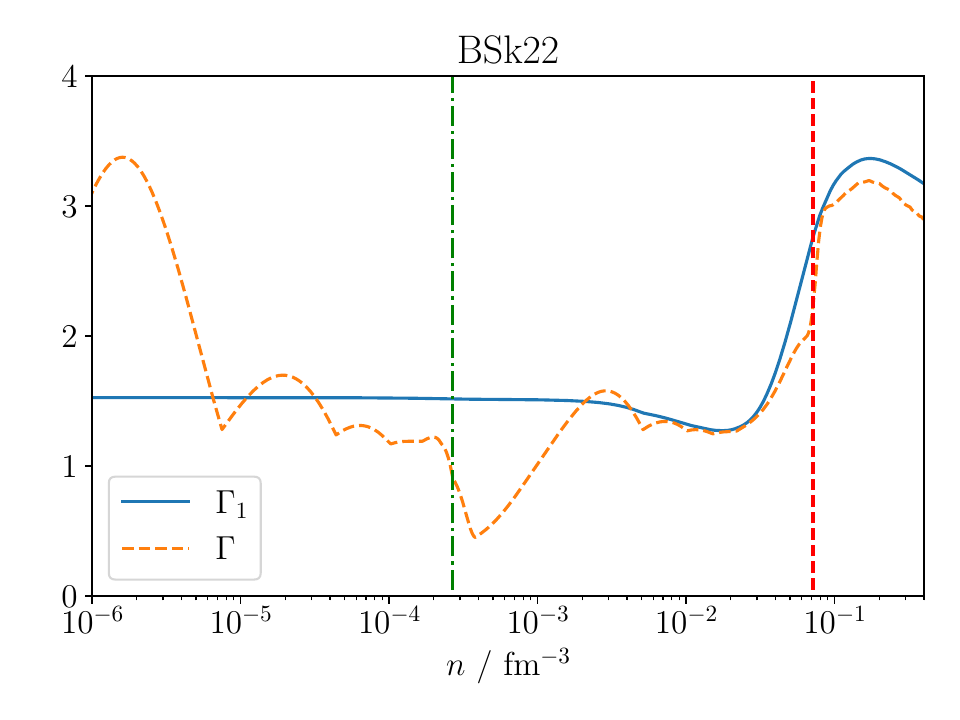}
    \caption{Plot of adiabatic and background indices $\Gamma_1$ and $\Gamma$ vs baryon number density $n$ for BSk22. The red dashed line corresponds to the location of the crust core interface density $n_{cc}$ and the green dot dashed line corresponds to the neutron drip point density $n_{nd}$ as given by \citet{BSkGR}.}
    \label{props}
\end{figure}

Moving on to the g-modes, we need to pay careful attention to the adiabatic index $\Gamma_1$. The results illustrated in Figure~\ref{BSkGam1s} highlight the behaviour in the neutron star core. In order to explore the g-modes we need to contrast the adiabatic index for the perturbations with that of the background, $\Gamma$, which is obviously taken to be in chemical equilibrium. We also need to extend the model to lower densities, beyond the core fluid. For the background model, this is done by using the fits from \citet{BSkGR} which smears out any sharp discontinuities, e.g. associated with distinct composition layers in the neutron star crust \citep{Crust1,Crust2}. These discontinuities would give rise to interface modes if not smoothed out, an issue discussed later on. The $\Gamma$ obtained for this model and the BSk22 example is shown in Figure \ref{props} (the plots for the other equations of state are very similar). The results show that the low-density background index varies significantly, a behaviour  inherited from the fits of \citet{BSkGR}. For the perturbations, we opt to simply extend the fit for $\Gamma_1$ from the core model to lower densities. This assumption is somewhat dubious, as it does not account for the  underlying microphysics in the crust (or indeed the elasticity of the nuclear lattice), but as is clear from Figure~\ref{props} it means that we are effectively treating $\Gamma_1$ as constant at low densities. This means that our low-density treatment is on a par with the vast majority of previous work on g-modes (see \citealt{Kokkotas,Kokkotas2} for recent examples) which assumes that the adiabatic index is constant throughout the star. In essence, our model may be inconsistent, but it is an improvement on previous work. 

Of course, we need to be mindful of the inconsistencies at low densities. Especially since there will be distinct mode features associated with this region. That this should be the case is evident from Figure~\ref{props}. It is generally the case that low-frequency waves (the g-modes) can propagate in regions where $\Gamma_1>\Gamma$. Figure~\ref{props} confirms that the neutron star core represents one such propagation region. In addition, we see that mode solutions may be supported in the low-density region near neutron drip. As we will soon see, this leads to the presence of a second family of g-modes, located in the neutron star crust.  The fact that our model is somewhat artificial at low densities means that these additional mode results must be considered with caution. However, that there should exist g-modes associated with the neutron star crust is known since the work by \citet{Reisenegger}. One would expect these modes to be sensitive to discontinuities between the different layers of nuclei \citep{FinnCrust, Miniutti}, a feature that is not present in our model. The dynamics of the crust region will also be sensitive to the associated elasticity, which is not included here. Future efforts should aim towards a consistent treatment of both core and crust, ideally making use of a consistent equation of state model that covers both regions. 

\begin{figure}
    \centering
    \includegraphics[height=6cm]{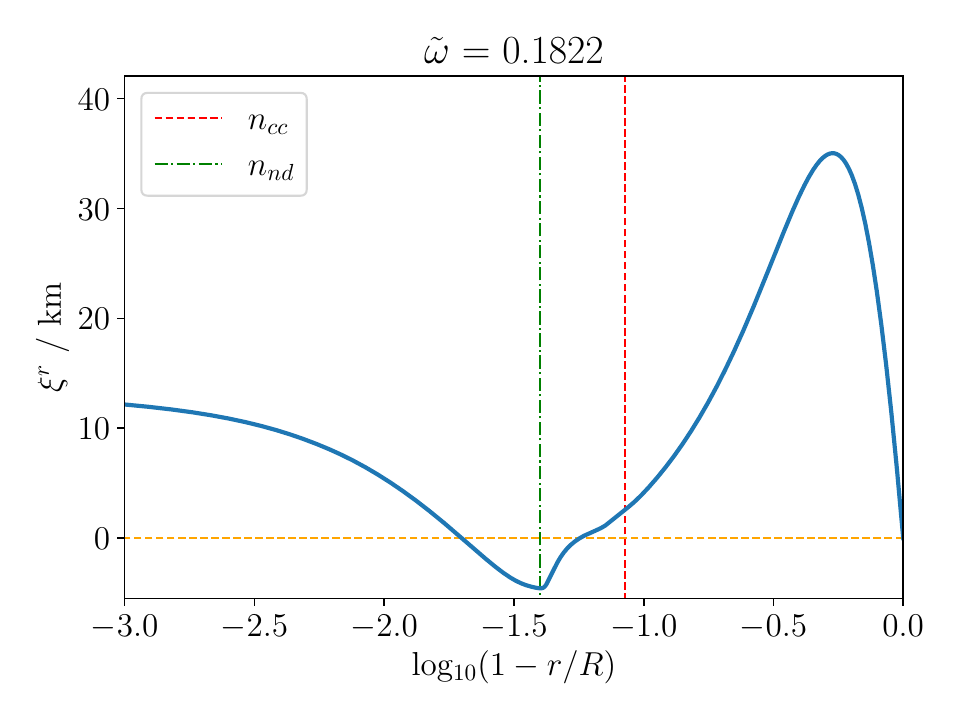}
    \includegraphics[height=6cm]{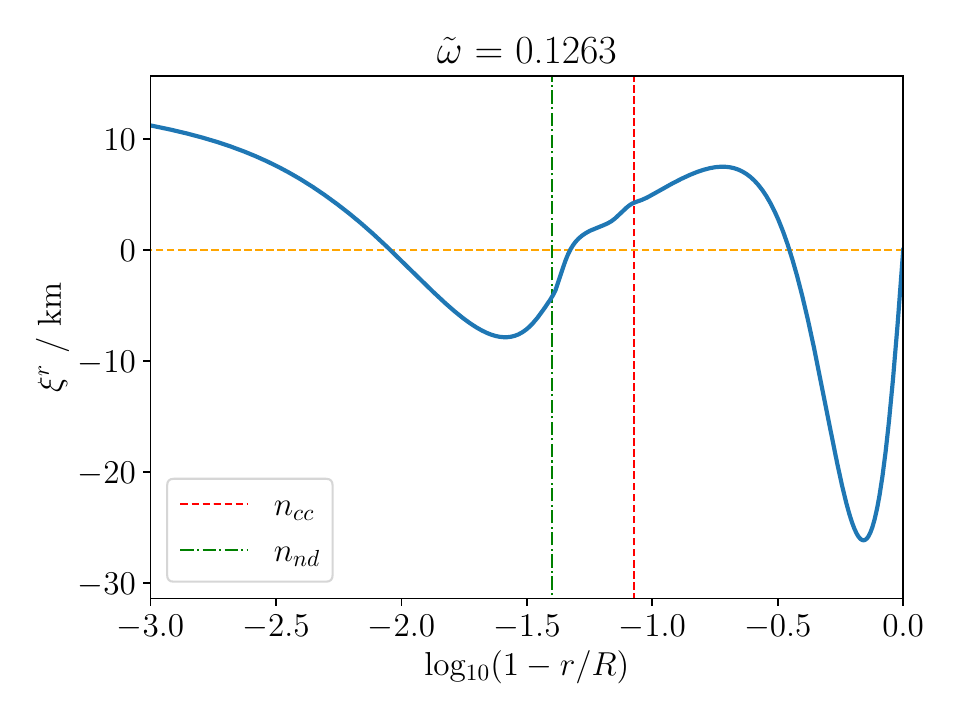}
    \caption{Plots of $\xi^r$ for the first two g-modes for a neutron star with $M=1.4 M_\odot$ and the BSk22 equation of state. On the left is $g_1$ and on the right $g_2$. The red dashed line corresponds to the location of the crust-core interface density $n_{cc}$ and the green dot-dashed line corresponds to the neutron drip point density $n_{nd}$ as given by \citet{BSkGR}. The orange line corresponds to $\xi^r=0$ and the dimensionless mode frequency \eqref{DimensionlessFreq} is given on top.}
    \label{BSk22Gmodes}
\end{figure}

\begin{figure}
    \centering
    \includegraphics[height=6cm]{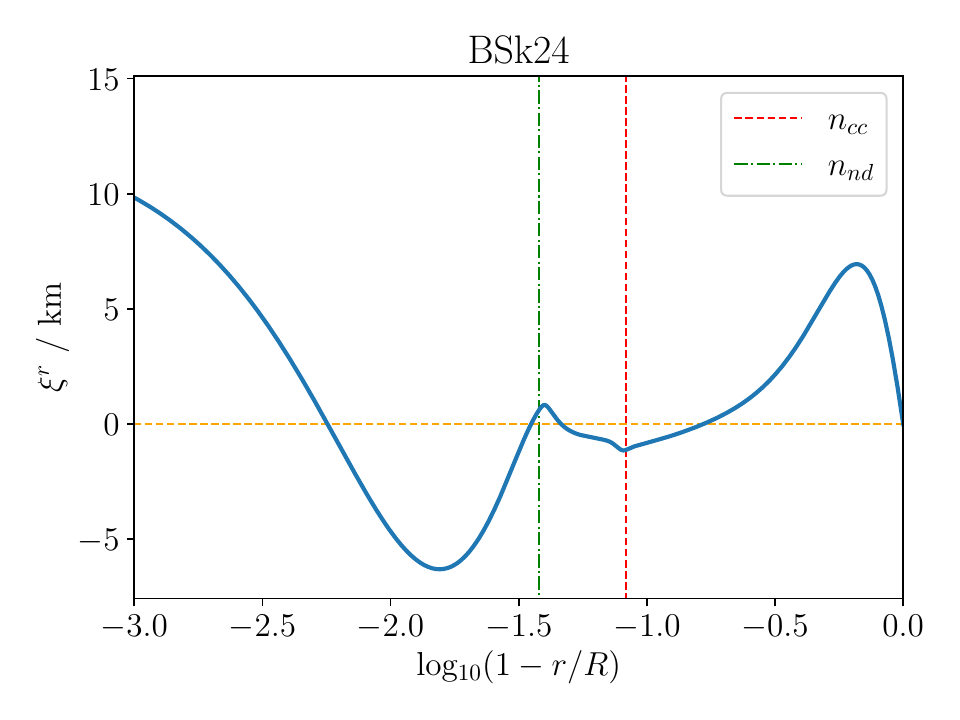}
    \includegraphics[height=6cm]{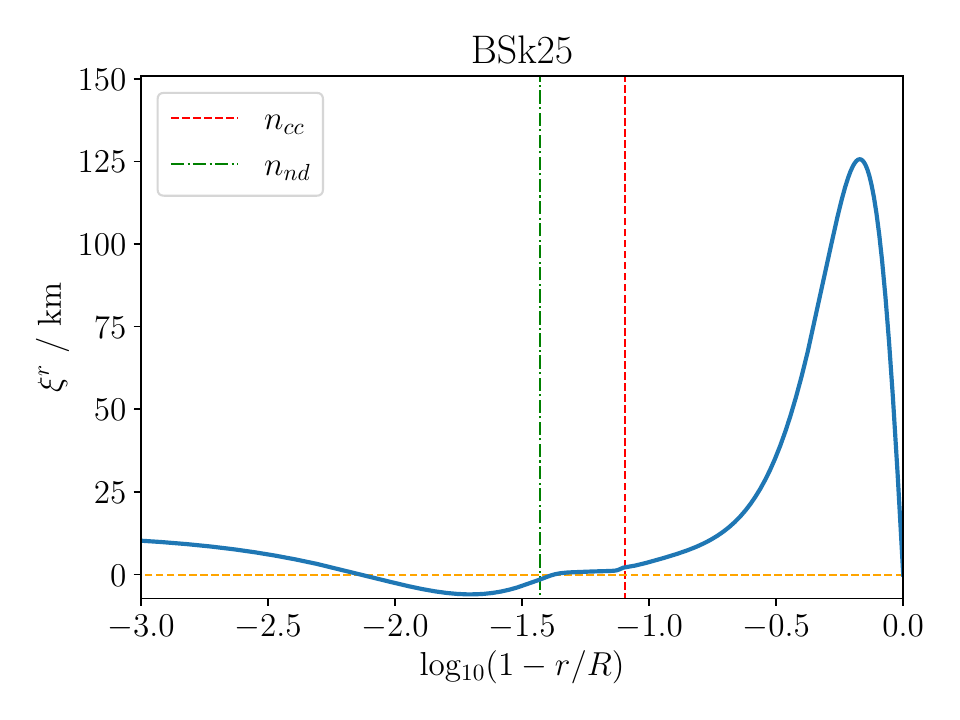}
    \includegraphics[height=6cm]{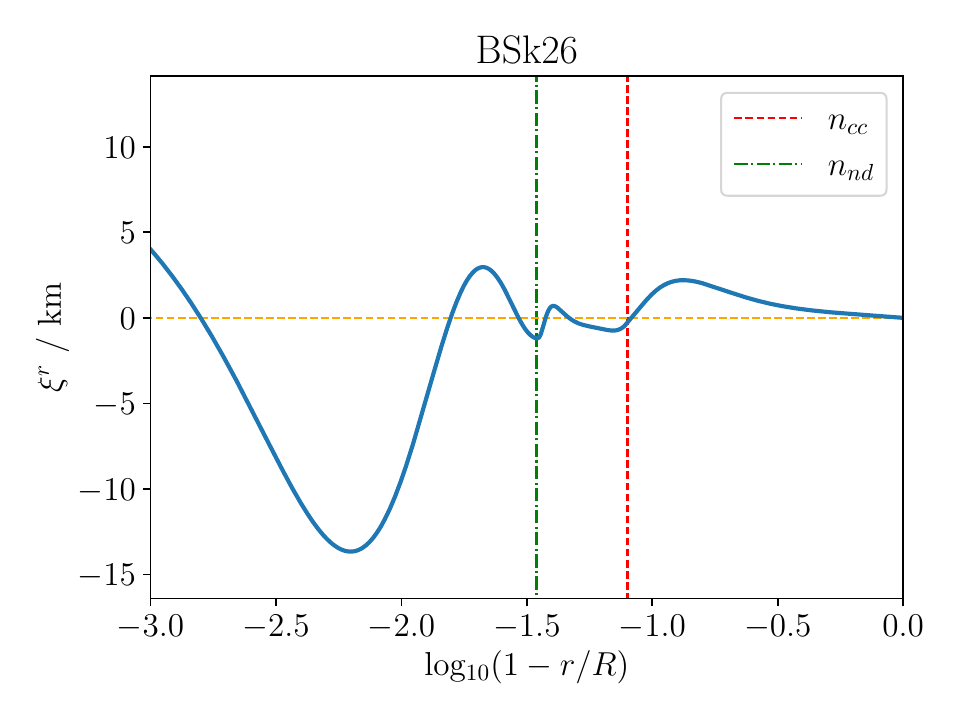}
    \caption{Plots of $\xi^r$ for the fundamental g-mode, $g_1$, for neutron stars with $M=1.4 M_\odot$ for the  BSk24-26  equations of state (as indicated in the respective panels). The red dashed line corresponds to the location of the crust-core interface density $n_{cc}$ and the green dot-dashed line corresponds to the neutron drip point density $n_{nd}$ for each equation of state, as given by \citet{BSkGR}. The orange line corresponds to $\xi^r=0$.}
    \label{GRG1Modes}
\end{figure}

The nature of the adiabatic and background indices is inherited by the g-mode solutions, which typically have a number of distinguishable features. Examples of this are provided in Figures~\ref{BSk22Gmodes} and \ref{GRG1Modes}. The first of these, Figure~\ref{BSk22Gmodes}, shows  the radial component of the fluid displacement $\xi^r$ vs $r$ for the fundamental g-mode, $g_1$,  and the first overtone, $g_2$ (which is identified by the presence of a node in the eigenfunction in the star's core), for the BSk22 model. Meanwhile, Figure~\ref{GRG1Modes} shows the fundamental g-modes for the other three equations of state, BSk24-26. All results are for stars with mass $M=1.4 M_\odot$.  In the figures, the red and green dashed lines correspond to the locations of the crust-core interface density, $n_{cc}$, and the neutron drip point density, $n_{nd}$, for each equation of state. These mark the boundaries in the equation of the state between the core, outer crust and inner crust. The values of the two densities were taken from \citet{BSkGR}. From the plots we note the characteristic g-mode amplitude peak in the core of the star and  additional features closer to the surface (for BSk26 the two peaks are notably of comparable magnitude). Different features of the eigenfunctions can be linked to  distinguishable features in the equation of state, like the sharp variation in the sound speed at neutron drip (see Figure~\ref{props}).  In fact, the behaviour near neutron drip resembles that expected for interface waves. However, in contrast to the results from \citet{gittins2024}, we are not dealing with distinct interface modes here.

Even though the four equation of state models belong to the same family, there are notable differences between the  mode solutions, such as how deep into the core the peak of the oscillation is located. 
Considering the electron fraction as a function of baryon number density from \citet{BSkGR}, we note that BSk22 has the highest electron fraction in the core while BSk26 has the lowest. In addition to  this, BSk26 has an additional peak at low density. Both  features help explain the behaviour for BSk26 seen in Figure~\ref{GRG1Modes}, i.e. the amplitude peak in the core being closer to the crust and being smaller  than for the other BSk models.

\begin{figure}
    \centering
    \includegraphics[height=6cm]{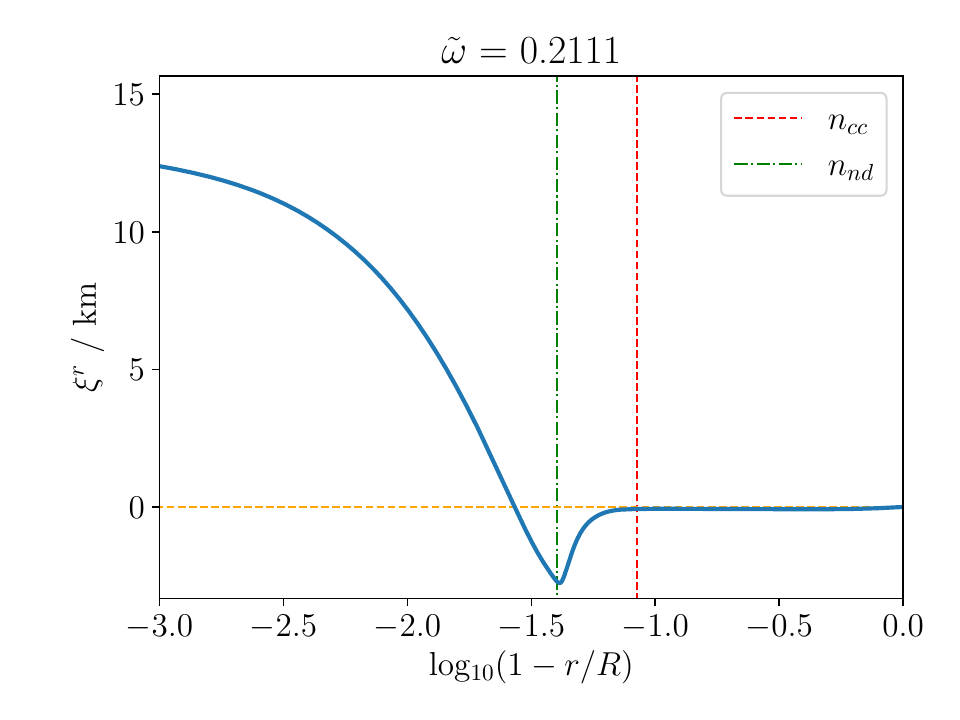}
    \includegraphics[height=6cm]{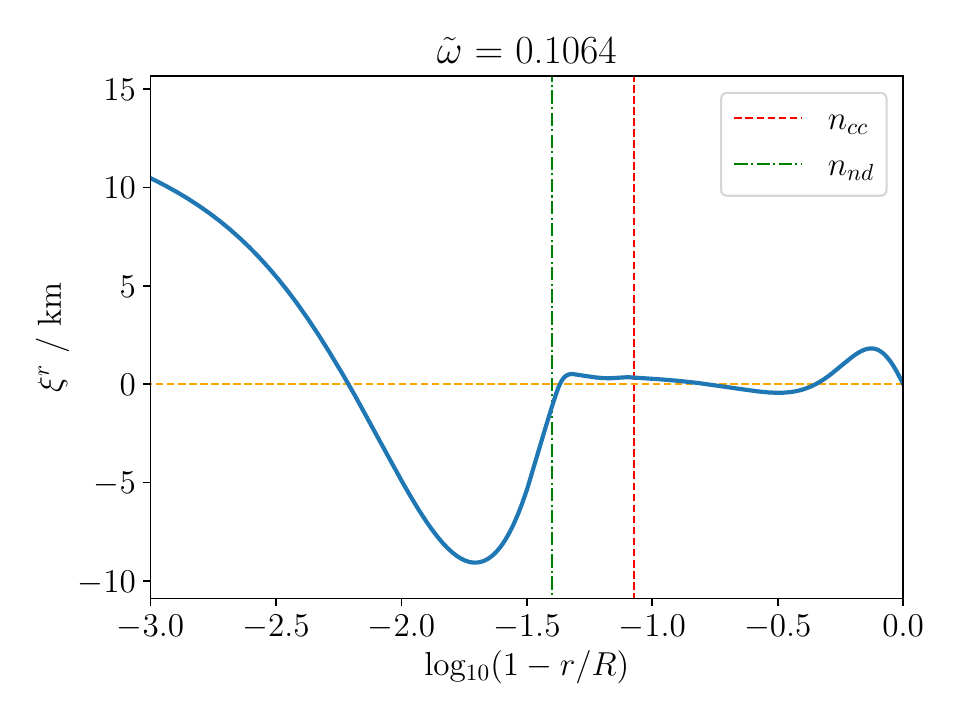}
    \caption{Plots of $\xi^r$ for the first two crustal g-modes for a neutron star with $M=1.4 M_\odot$ and the BSk22 equation of state. On the left is the first crustal mode and on the right is the second one. The red dashed line corresponds to the location of the crust-core interface density $n_{cc}$ and the green dot-dashed line corresponds to the neutron drip point density $n_{nd}$ as given by \citet{BSkGR}. The orange line corresponds to $\xi^r=0$ and the dimensionless mode frequency \eqref{DimensionlessFreq} is given on top.}
    \label{CrustModes}
\end{figure}

As already suggested,  the presence of the crustal region leads to the appearance of an additional set of modes.   Examples of these solutions are shown in Figure~\ref{CrustModes} for BSk22, again for a star with $M=1.4M_\odot$. These modes are clearly identified by a more significant fluid motion in the crust region, while the mode amplitude in the star's core remains low. This allows us to distinguish the two families of mode solutions. In order to confirm that the second set of modes owe their existence to the low-density propagation region identified in Figure~\ref{props} we have also calculated the modes for a model where we set $\Gamma_1=\Gamma$ at densities below neutron drip (adopting the strategy from \citealt{HotGmodes}). The results of this exercise show that the first core g-modes are not significantly affected, while the crust g-modes disappear from the spectrum, see Table~\ref{CrustTable}. This accords with our expectations.

\begin{table}
\small
\centering
\caption{The dimensionless mode frequencies \eqref{DimensionlessFreq} for the first two $l=2$ core and crust g-modes using the BSk22 equation of state for a few chosen gravitational masses. The results were obtained from the full analytic $\Gamma_1$ and also setting $\Gamma_1=\Gamma$ in the crust while retaining the original $\Gamma_1$ in the core. The results confirm that one set of the identified $g$-modes originates from the physics in the neutron star crust. }

\begin{tabular}{||c|c|c|c|c||}
\hline
\multicolumn{1}{||c|}{\multirow{2}{*}{$M_\odot$}} & \multicolumn{2}{|c|}{\multirow{2}{*}{Mode}}& \multicolumn{1}{|c|}{Full $\Gamma_1$}& \multicolumn{1}{|c|}{$\Gamma_1=\Gamma$ in crust} \\
\cline{4-5}
\multicolumn{1}{||c|}{} & \multicolumn{1}{|c|}{} & \multicolumn{1}{|c|}{} & $\tilde\omega_n$ & $\tilde\omega_n$\\
\hline
\multirow{6}{*}{}  
1.4	&	core	&	$g_1$	&	0.1822	&	0.1814	\\
	&		&	$g_2$	&	0.1263	&	0.1219	\\
	&	crust	&	$g_1$	&	0.2111	&	-	\\
	&		&	$g_2$	&	0.1064	&	-	\\
1.6	&	core	&	$g_1$	&	0.1832	&	0.1831	\\
	&		&	$g_2$	&	0.1240	&	0.1221	\\
	&	crust	&	$g_1$	&	0.1890	&	-	\\
	&		&	$g_2$	&	0.0934	&	-	\\
1.8	&	core	&	$g_1$	&	0.1850	&	0.1847	\\
	&		&	$g_2$	&	0.1232	&	0.1223	\\
	&	crust	&	$g_1$	&	0.1689	&	-	\\
	&		&	$g_2$	&	0.0847	&	-	\\
2	&	core	&	$g_1$	&	0.1860	&	0.1858	\\
	&		&	$g_2$	&	0.1224	&	0.1220	\\
	&	crust	&	$g_1$	&	0.1501	&	-	\\
	&		&	$g_2$	&	0.0745	&	-	\\
            \hline
\hline
\end{tabular}
\label{CrustTable}
\end{table}

The results in Table~\ref{CrustTable} also show that the core/crust g-mode frequencies are interleaved in the mode spectrum. The table provides results for the dimensionless  frequencies, based on the scaling
\begin{equation}
    \tilde \omega_n^2 = \frac{\omega_n^2}{GM/R^3}.
    \label{DimensionlessFreq}
\end{equation}

\begin{figure}
    \centering
    \includegraphics[height=5.9cm]{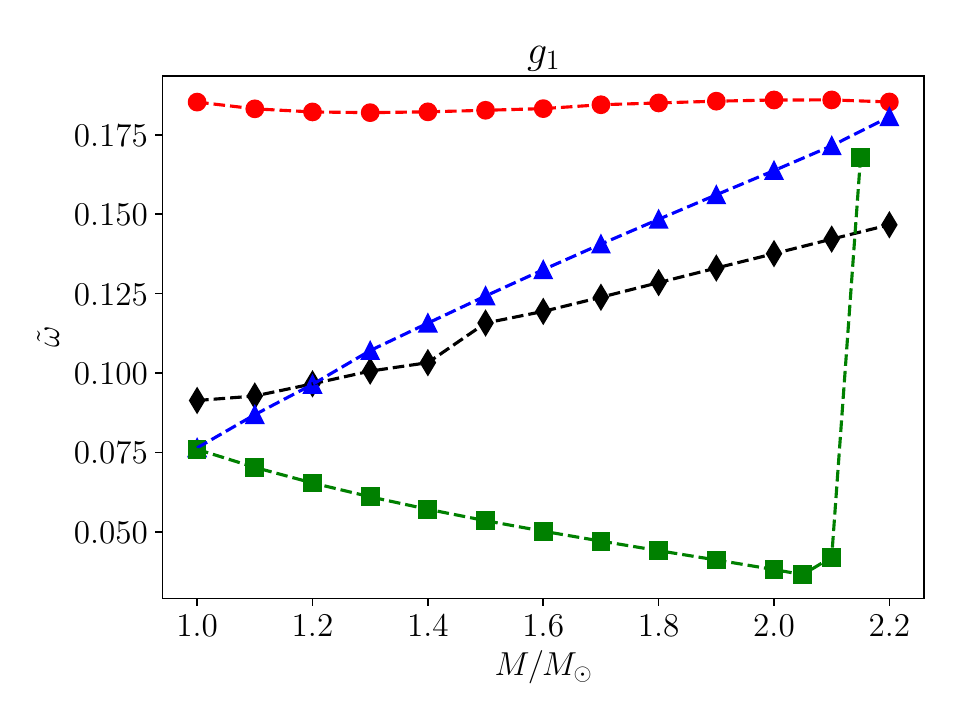}
    \includegraphics[height=5.9cm]{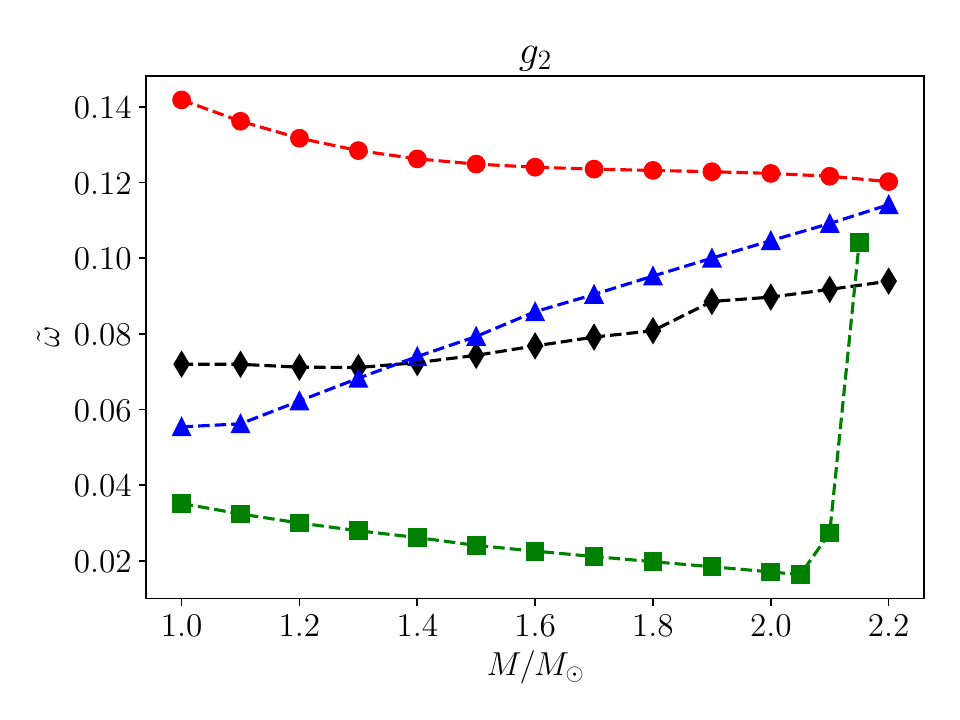}
    \includegraphics[height=5.9cm]{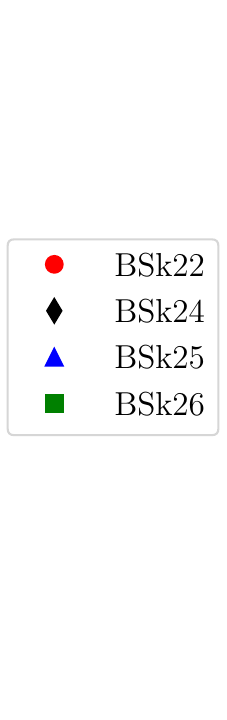}
    \caption{Plots of dimensionless frequencies $\tilde \omega$ \eqref{DimensionlessFreq} vs total mass energy $M/M_\odot$ for the g-modes $g_1$ and $g_2$ for the BSk22-26 equations of state. On the left is the plot for the fundamental g-mode, $g_1$, and on the right is the plot for its first overtone $g_2$.}
    \label{BSkG1and2}
\end{figure}

Moving on, we examine the dependence of the g-modes on the  stellar mass $M$.  Results for the first two core $g$-modes are shown in Figure \ref{BSkG1and2},  plotting $\tilde \omega$ vs $M$ for the different BSk equations of state. The range of stellar masses was chosen to be from $1M_\odot$ up to the approximate maximum masses given by \citet{BSkGR}. Comparing to the f-modes from Figure \ref{fmodeuniversal}, we note that there is much more variance in the behaviour between the different equations of state. For BSk22 and BSk26, $\tilde \omega$ tends to decrease with increasing $M$, whereas the opposite behaviour is seen for BSk24 and BSk25. We also note a sharp change in the results near the maximum mass for BSk26. 

\begin{figure}
    \centering
    \includegraphics[height=6cm]{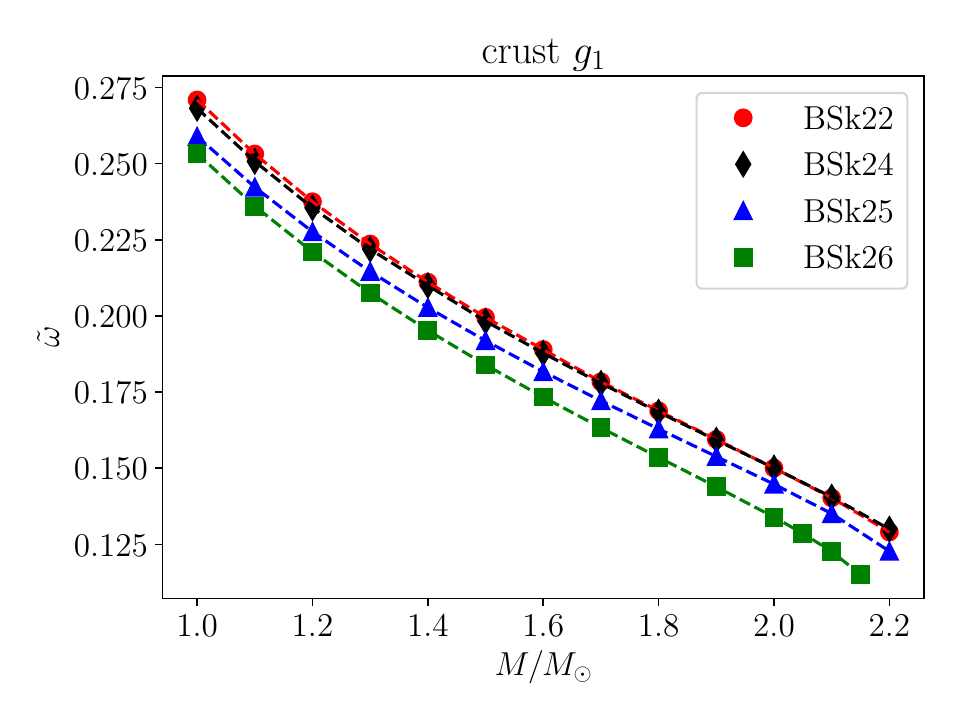}
    \caption{Plot of dimensionless frequency $\tilde \omega$ \eqref{DimensionlessFreq} vs total mass energy $M/M_\odot$ for the first crustal g-mode for the BSk22-26 equations of state.}
    \label{CrustHz}
\end{figure}

Finally, returning to the crustal g-modes, their frequency varies slightly with the mass again decreasing with increasing $M$. This is evident from Figure \ref{CrustHz}, and should be  expected as $n_{cc}$ and $n_{nd}$ do not depend on $M$, therefore the fraction of the stellar mass that is the crust, decreases with increasing $M$. This result agrees with the findings of \citet{Reisenegger}. Unlike the core g-modes, there is minimal difference between the crustal modes of the different BSk models. Again this is to be expected as they depend more on the crustal parameters than $\Gamma_1$.

\subsection{Detectability of tidal resonances}

Having determined the g-mode solutions we are well placed to discuss to what extent the modes are detectable, e.g. through the associated tidal resonance in a binary inspiral signal. For a binary neutron star system---or, indeed, a black hole-neutron star system---a resonance occurs between the mode frequency  and the orbital frequency when (for quadrupole modes with $m=2$) $\omega_n \approx 2\Omega_{orb}$. This causes energy to be transferred into the mode, exciting the oscillation and drawing energy from the orbit \citep{DongLai,WynnAstereo}. As the orbit loses energy, it shrinks faster, manifesting as a change in the gravitational waveform \citep{WaveformConstrain}. The goal is then to quantify this shift in the waveform for a given mode. The core g-modes are particularly interesting in this respect because, even though they are expected to leave a weak signature, detection would help constrain the composition and state of matter in the neutron star interior. Previous work suggests that such detections may, just about, be within the reach of next-generation gravitational-wave instruments \citep{Wynn2}.

The problem of tidal resonances is not yet completely formulated in general relativity, so we will resort to a hybrid approximation based on the Newtonian tidal interaction.
First, let us assume a Newtonian orbit, with masses $M_1$ and $M_2$, an orbital separation $D(t)$ and orbital frequency $\Omega_{orb}$. Then, the shift in orbital phase $\Delta \Phi$ due to energy transfer $\Delta E$ during the inspiral can be estimated as \citet{DongLai}
\begin{equation}
    \frac{\Delta \Phi}{2\pi}\approx -\frac{t_D}{t_{orb}}\frac{\Delta E}{|E_{orb}|},
\end{equation}
where $E_{orb}$ is the orbital energy given by
\begin{equation}
    E_{orb}=-\frac{GM_1M_2}{2D},
\end{equation}
 $t_{orb}$ is the orbital period, given by
 \begin{equation}
     t_{orb}=\frac{2\pi}{\Omega_{orb}},
 \end{equation}
 and $t_D$ is the orbital decay timescale given by
 \begin{equation}
     t_D=\frac{D}{|\dot D|},
 \end{equation}
where the dot indicates a derivative with respect to time. To leading order, the orbital separation evolves due to the emission of gravitational waves as 
\begin{equation}
    \dot D = - \frac{64}{5c^5}\frac{M_1M_2(M_1+M_2)}{D^3}.
\end{equation}
The mode resonance then occurs when the mode frequency $\omega_n$ is twice the orbital frequency. This can  be related to gravitational wave frequency $f$ by
\begin{equation}
    \omega_n=2\Omega_{orb}=2\pi f.
\end{equation}
One can also relate $\Omega_{orb}$ to $D$ by Kepler's law
\begin{equation}
    \Omega_{orb}=\sqrt{\frac{G(M_1+M_2)}{D^3}}.
\end{equation}
For simplicity, it is assumed from now on that the two masses are equal, $M=M_1=M_2$. 

The last piece of the puzzle involves quantifying $\Delta E$. From \citet{DongLai}, one can relate the energy transfer to the mode overlap integral $Q_{nl}$, the parameter quantifying the coupling between the tide and the mode,%
\footnote{%
    For  comparison with its relativistic analogue \eqref{relQnl}, the overlap integral in Newtonian gravity is defined as
    \begin{equation*}
        Q_{nl} = \int \rho\xi^{i*}_n\nabla\left(r^l Y_{lm}\right)\dd V,
    \end{equation*}
    where $\rho$ is the mass density.
}
as%
\begin{equation}
    \Delta E \approx \frac{\pi^2}{512}\frac{GM^2}{R}\omega_n^{1/3}Q^2_{nl}\left(\frac{Rc^2}{GM}\right)^{5/2}.
\end{equation}
Putting all this together, we have
\begin{equation}
    \frac{\Delta \Phi}{2\pi}\approx -\frac{5 \pi}{4096}\left(\frac{c^2 R}{GM}\right)^5\frac{\tilde Q_{n l}^2}{\tilde \omega^2_{nl}},
    \label{ShiftAndQ}
\end{equation}
where $\tilde \omega_n$ is the dimensionless mode frequency \eqref{DimensionlessFreq} and we also normalised the overlap integral to $MR^l$ \citep{Wynn2}. We will assume that a similar expression applies in the relativistic case---although this remains to be justified by a detailed derivation---with the scaling based on the fully relativistic results for the star's mass and radius.

In order to work out the overlap integral from the Cowling mode results, we consider a binary neutron star system, where the orbit is in the equatorial plane and it is assumed each star sees the other as a point mass. 
The gravitational energy that one of the  stars absorbs from the other is given by the tidal part of the Hamiltonian,
\begin{equation}
    H_{tid} = \int \Phi^T \frac{\delta \varepsilon^*}{c^2} \sqrt{-g}d^3x,
    \label{TidalEnergy}
\end{equation}
where $\Phi^T$ is the tidal potential experienced by the star and the integral is over the star's volume \citep{Kokkotas}.

In order to assess the relevance of the  low-frequency g-mode resonances, we use the Newtonian result to approximate the tidal potential as in \citet{Kokkotas}.  From \cite{2024MNRAS.527.8409P} we then have
\begin{equation}
    \Phi^T=-GM\sum_{l=2}^\infty \sum_{m=-l}^l \frac{W_{lm}r^l}{D(t)^{l+1}}Y_{lm}e^{-im\psi},
    \label{TidalPotential}
\end{equation}
where $\psi$ is the orbital phase and  $W_{lm}=0$ for odd $l+m$, else
\begin{equation}
    W_{lm}=(-1)^{(l+m)/2}\sqrt{\frac{4\pi}{2l+1}(l-m)!(l+m)!}\left[2^l\left(\frac{l+m}{2}!\right)\left(\frac{l-m}{2}!\right)\right].
\end{equation}
Now consider a motion of the neutron star $\xi^i$, it can be expressed as the sum over modes,
\begin{equation}
    \xi^i=\sum_{nlm}q_{nlm}\xi^i_{nlm} e^{i\omega_{nlm} t},
    \label{modexis}
\end{equation}
where $q_{nlm}$ is the amplitude of each mode. Here the eigenfunctions $\xi^i_{nlm}$ satisfy the relevant  mode equations, which in this work are taken to be \eqref{Z1} and \eqref{Z2}. 
Now, given the mode orthogonality, $\delta \varepsilon$ can be expressed as the sum of contributions of all the modes, 
\begin{equation}
    \delta\varepsilon = \sum_{nlm} \delta\varepsilon_{nlm} e^{i\omega_{nlm}t}.
    \label{epsilonmode}
\end{equation}
From \eqref{delepsilon} we know that
    \begin{equation}
  \partial_t \delta \varepsilon + \partial_t \xi^i \partial_i  \varepsilon  + \frac{(p+\varepsilon)  e^{\nu/2}}{ \sqrt{-g}} \partial_i \left[ e^{-\nu/2}\sqrt{-g} \partial_t \xi^i \right] = 0.
  \label{epsilonalpha}
\end{equation}
Using \eqref{modexis} and \eqref{epsilonmode} gives
\begin{equation}
    \delta \varepsilon_{nlm} = -q_{nlm} \left[ \xi^i_{nlm} \partial_i \varepsilon +\frac{(p+\varepsilon)  e^{\nu/2}}{ \sqrt{-g}} \partial_i \left( e^{-\nu/2}\sqrt{-g} \xi^i_{nlm}. \right)\right].
\end{equation}
This can then be put into \eqref{TidalEnergy} and, after simplifying and integrating by parts, we arrive at
\begin{equation}
    H_{tid}=-\sum_{nlm} q^*_{nlm} \int \frac{\varepsilon+p}{c^2}\xi^{i*}_{nlm} \nabla_i(\Phi^T)\sqrt{-g} d^3x.
\end{equation}
Finally, using \eqref{TidalPotential},
\begin{equation}
    H_{tid}=-GM\sum_n \sum_{l=2}^\infty \sum_{m=-l}^l q^*_{nlm}\frac{W_{lm}}{D(t)^{l+1}}Q_{n l}e^{-im\psi}
\end{equation}
where $Q_{n  l}$ is identified as the relativistic analogue of the tidal overlap integral, given by
\begin{equation}
    Q_{n l}=\int\frac{\varepsilon+p}{c^2}\xi^{i*}_{n}\nabla_{i}(r^lY_{lm})\sqrt{-g}d^3x,
    \label{relQnl}
\end{equation}
which agrees with  the expression used by \citet{Kokkotas}. Substituting in \eqref{static xi} from the previous section, for a specific mode, the dimensionless overlap integral simplifies to
\begin{equation}
    \tilde Q_{n l}=\frac{1}{M R^l}\int_{0}^{R}e^{(\nu+\lambda)/2}\frac{\varepsilon+p}{c^2}r^l\left[lW_l+l(l+1)V_l\right]d r.
    \label{CowlingQTilde}
\end{equation}
Note that, as the star is not  rotating, there is no explicit dependence on $m$.

\begin{figure}
    \centering
    \includegraphics[height=6cm]{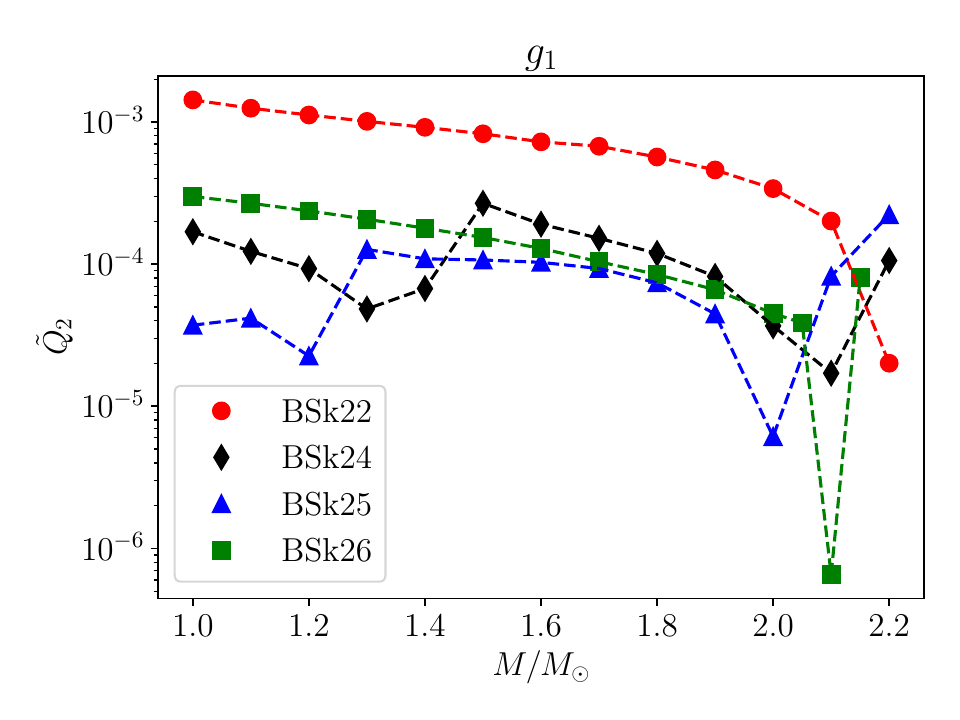}
    \includegraphics[height=6cm]{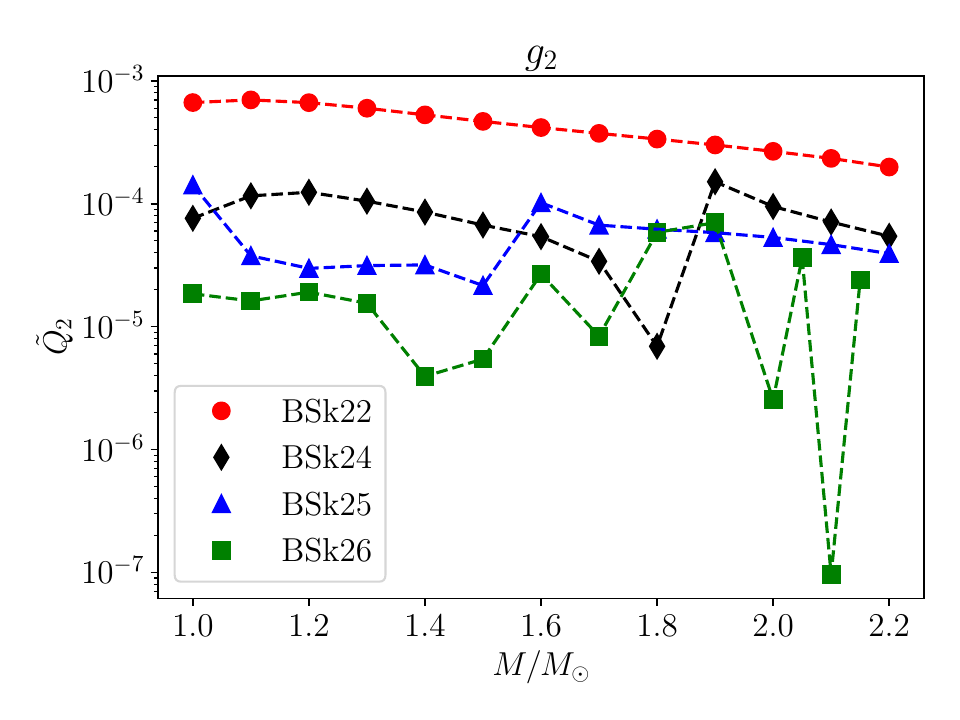}
    \includegraphics[height=6cm]{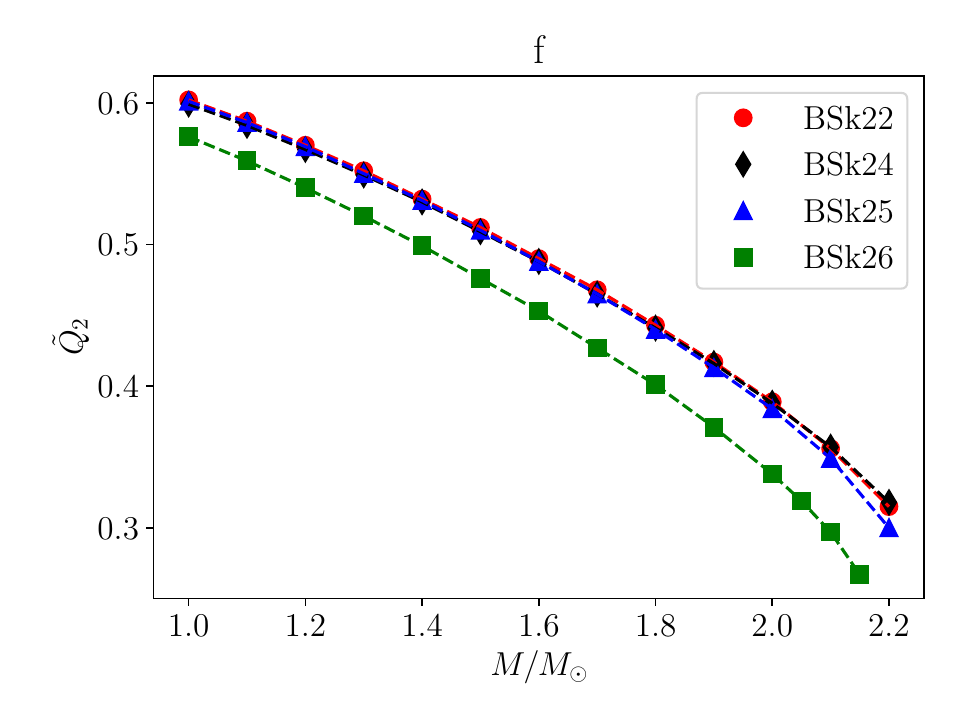}
    \caption{Plot of dimensionless overlap integral $\tilde Q_{n 2}$ \eqref{CowlingQTilde} vs total mass energy $M$ for the fundamental f-mode and the first two core g-modes for the BSk22-26 equations of state.}
    \label{QsBSk}
\end{figure}

We have calculated the overlap integral $\tilde Q_{n 2}$  for the f-mode and the first 2 core g-modes for a range of neutron star masses and the four different BSk equations of state (and equal mass systems). For the calculation of the overlap integrals, we will adopt the more common normalisation choice $\mathcal{A}^2=MR^2$, as defined in \eqref{A2Norm}, instead of that of \eqref{XiNorm}. The sample of  results provided in Table \ref{GRQTable} give useful insight into how the tidal coupling varies with the stellar parameters. 
From Figure \ref{QsBSk} it is immediately clear that the value of $\tilde Q_{n 2}$ for the f-mode is several orders of magnitude greater than for the g-modes. This is expected due to the f-mode most closely resembling the tidal potential, and obviously accords with the Newtonian results. A closer inspection of Figure \ref{QsBSk} shows that for f-modes, the results for the different equations of state are very similar to one another. In fact, with the exception of BSk26, the results are almost indistinguishable. For the g-modes the behaviour is notably less regular. This is to be  expected, given that the g-modes are more sensitive to local variations at different densities inside the star. Still, in general we note that $\tilde Q_{n 2}$ tends to decrease with the mass of the star, as one would expect from \eqref{CowlingQTilde}. 

\begin{figure}
    \centering
    \includegraphics[height=6cm]{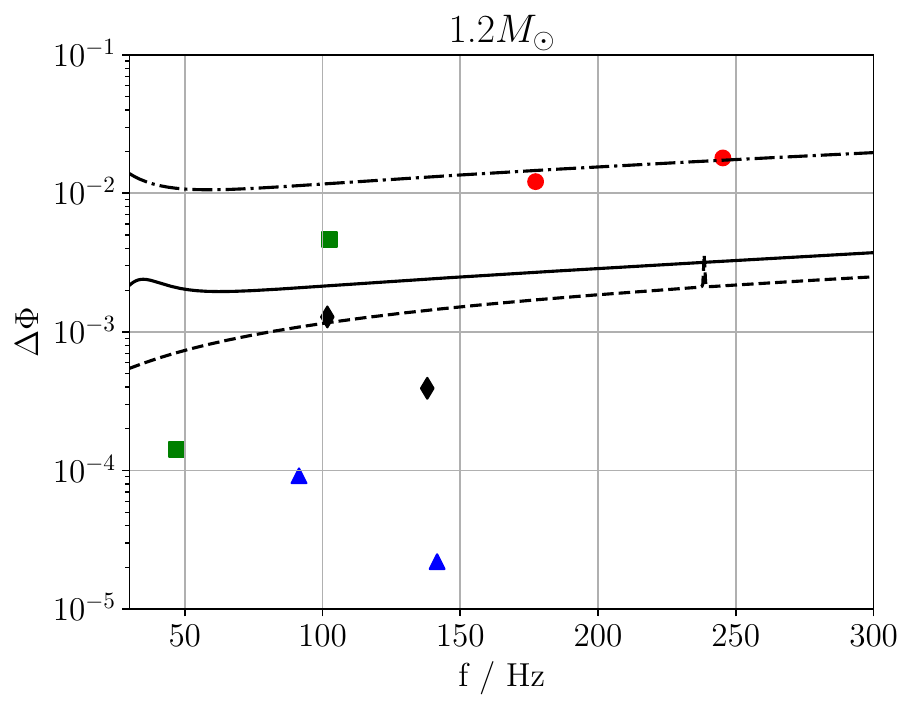}
    \includegraphics[height=6cm]{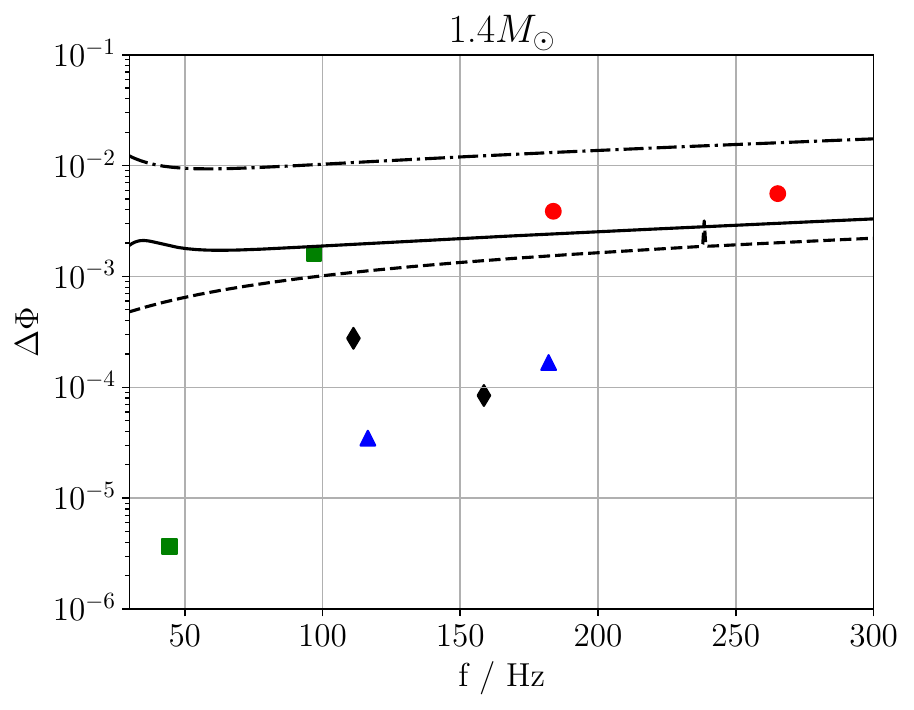}
    \includegraphics[height=6cm]{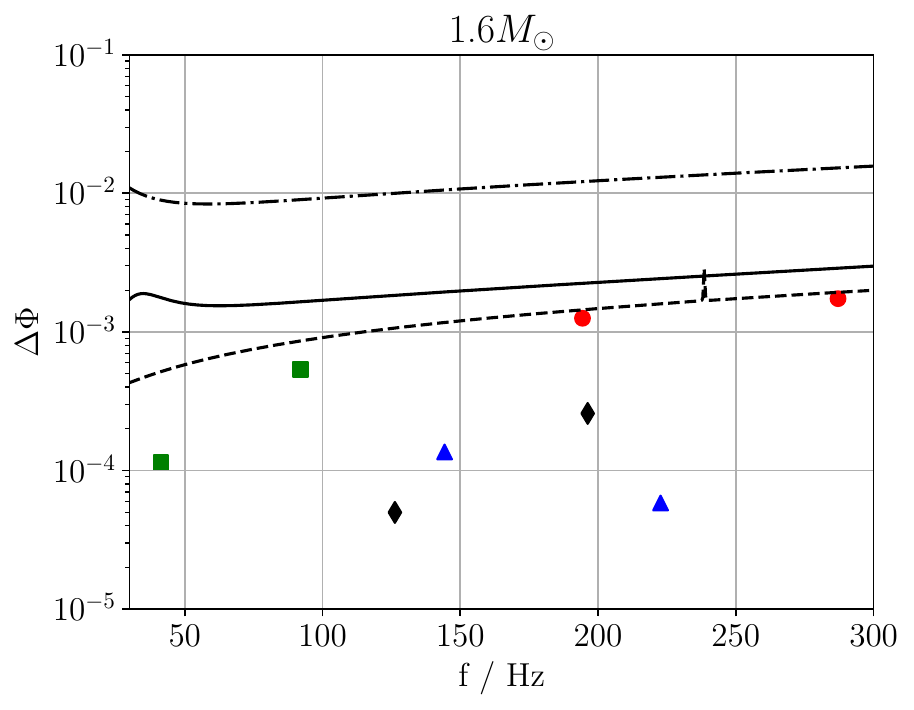}
    \includegraphics[height=6cm]{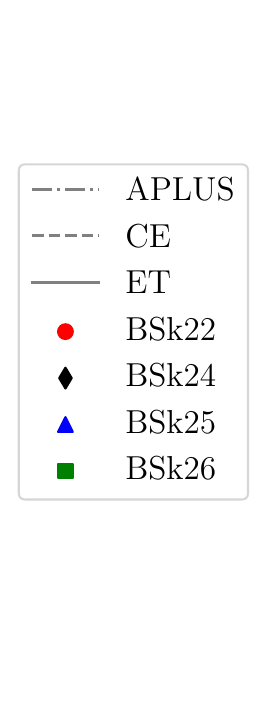}
    \caption{Plot of detectable shift in orbital phase  $\Delta \Phi(f)$ \eqref{ShiftAndQ} vs frequency for the first two core g-modes for the BSk22-26 equations of state. The curves are upper limits calculated using $\Delta \Phi$ from \eqref{DetectorSensitivity} for LIGO A+ (long-dashed), the Einstein Telescope (ET; solid) and Cosmic Explorer (CE; short-dashed). Each plot is labelled by the mass energy $M$ of each individual neutron star. All the binaries are assumed to be equal mass systems and located 40Mpc from the detectors.}
    \label{SensitivityQ}
\end{figure}

Moving on to the question of detectability, we follow the arguments from \citet{Wynn2}--- combining \eqref{ShiftAndQ} with the results from Table \ref{GRQTable}. The inferred value for the change in phase associated with each resonance is plotted against frequency in Figure \ref{SensitivityQ} alongside the sensitivity curves of LIGO A+, Cosmic Explorer (CE) and the Einstein Telescope (ET), at a distance of 40 Mpc (the inferred distance to GW170817). In order to generate the curves for each detector we need to quantify the detectable shift in orbital phase as a function of frequency $\Delta \Phi(f)$ for each one. This is calculated from
\begin{equation}
    |\Delta \Phi|=\frac{\sqrt{S_n(f)}}{2 A(f) \sqrt{f}},
    \label{DetectorSensitivity}
\end{equation}
where $S_n(f)$ is the power spectral density of the noise for each detector and $A(f)$ is the gravitational wave amplitude of a generated test waveform as discussed by \citet{Read_2023}.
One still then needs to choose a waveform model. For simplicity, noting that recent work by \citet{Read_2023} shows that the differences in current waveform models are much smaller than the estimated data uncertainties for the low frequencies we are interested in, only one waveform model was considered. If a mode is detectable for one waveform model it should be detectable for others. Specifically, we used the IMRPhenomPv2$\_$NRTidal model as it includes the  static tidal deformabilities \citep{InformModel}. It is important to note that, as the inferred phase uncertainties are estimated from expected gravitational-wave data and existing waveform models, they only serve as approximate upper limits on any effects  
 not considered in the waveform modelling, like the influence of dynamical tides considered here. 

Our results are shown in Figure \ref{SensitivityQ}   with each panel labelled with the total mass $M$ of each neutron star in the binary.  We see that, in general, low mass stars are the most promising for detection for these g-modes. This is not surprising because of  the $M^{-5}$ dependence in \eqref{ShiftAndQ}. As expected from Figure \ref{QsBSk}, BSk22 appears most promising while the other three equations of state perform similarly.


\begin{table}
\normalsize
\centering
\caption{Dimensionless mode frequencies \eqref{DimensionlessFreq} for different $l=2$ core g-modes, as well as for the f-mode, for the four BSk equations of state and total mass energies $1.2M_\odot$, $1.4M_\odot$ and $1.6M_\odot$. Also calculated is the dimensionless overlap integral $Q_{n2}$ for each mode using \eqref{CowlingQTilde}. This data was used to produce Figure~\ref{SensitivityQ}. The notation e-p at the end of each number stands for $\times 10^{-p}$.}

\begin{tabular}{||c|c|c|c|c|c|c|c|c|c|}
\hline
\multicolumn{1}{||c|}{\multirow{2}{*}{$M_\odot$}} & \multicolumn{1}{|c|}{\multirow{2}{*}{Mode}}& \multicolumn{2}{|c|}{BSk22}& \multicolumn{2}{|c|}{BSk24}& \multicolumn{2}{|c|}{BSk25}& \multicolumn{2}{|c|}{BSk26} \\
\cline{3-10}
\multicolumn{1}{||c|}{} & \multicolumn{1}{|c|}{} & $\tilde\omega_n$ & $\tilde Q_{n2}$ & $\tilde\omega_n$ & $\tilde Q_{n2}$ & $\tilde\omega_n$ & $\tilde Q_{n2}$ & $\tilde\omega_n$ & $\tilde Q_{n2}$\\
\hline
\multirow{6}{*}{}  
                            1.2&$f$ &1.4996	&	5.70e-1	&	1.4885	&	5.67e-1	&	1.4801	&	5.69e-1	&	1.4730	&	5.40e-1\\
                               &$g_1$&0.1822	&	1.12e-3	&	0.0966	&	9.30e-5	&	0.0964	&	2.26e-5	&	0.0654	&	2.37e-4\\
                               &$g_2$&0.1317	&	6.65e-4	&	0.0712	&	1.24e-4	&	0.0623	&	2.98e-5	&	0.0300	&	1.91e-5\\
                               \hline1.4&$f$ &1.4353	&	5.32e-1	&	1.4224	&	5.30e-1	&	1.4140	&	5.31e-1	&	1.4028	&	4.99e-1\\
                                &$g_1$&0.1822	&	9.17e-4	&	0.1032	&	6.74e-5	&	0.1157	&	1.09e-4	&	0.0571	&	1.79e-4\\
                                &$g_2$&0.1263	&	5.29e-4	&	0.0724	&	8.56e-5	&	0.0740	&	3.18e-5	&	0.0262	&	3.94e-6\\
                               \hline1.6&$f$&1.3766	&	4.90e-1	&	1.3628	&	4.88e-1	&	1.3544	&	4.88e-1	&	1.3372	&	4.53e-1\\
                                &$g_1$&0.1832	&	7.25e-4	&	0.1194	&	1.91e-4	&	0.1325	&	1.03e-4	&	0.0502	&	1.29e-4\\
                                &$g_2$&0.1240	&	4.17e-4	&	0.0768	&	5.40e-5	&	0.0859	&	1.02e-4	&	0.0226	&	2.68e-5\\
            \hline
\hline
\end{tabular}
\label{GRQTable}
\end{table}

\section{Conclusions}\label{Conclusions}

By employing four models from the BSk equation of state family, we have shown how  subtle differences in the nuclear matter assumptions impact on the g-mode oscillation spectrum of neutron stars. This was achieved by setting up the linear perturbation equations in a dimensionless formalism, within the Cowling approximation, in the spirit of work by \citet{McDermott} and others. The advantage of the BSk family of models is that it allows us to employ a realistic description of the matter stratification, which is required  to calculate the g-mode spectrum. 

Along with the expected core g-modes, another class of mode solutions was observed, determined to be crustal g-modes of the kind discussed by \citet{Reisenegger}. Such modes arise due to distinct features in the equation of state, like the crust-core interface and the onset of neutron drip, both of which are encoded into the pressure and energy density functionals of \citet{BSkGR} that we employ in our calculations. These modes are, however, sensitive to aspects of the low-density physics (like the crust elasticity) that are not included in our model and hence the corresponding results are not expected to be robust. The fact that the modes originate in the crust is demonstrated by the modes disappearing from the oscillation spectrum when we artificially set $\Gamma_1=\Gamma$ in the low-density region. This exercise also demonstrates that the core g-modes, which provide the main focus for our discussion, are fairly insensitive to the low-density physics.

Once the mode frequencies were obtained, we examined the possibility of detecting these modes with next generation ground-based detectors such as Cosmic Explorer \citep{CosmicExplorer} and the Einstein Telescope \citep{ET}. This was achieved by calculating the overlap integral $Q_{n,2}$ for the $l=2$ modes and relating this to the detectable shift in orbital phase as a function of frequency $\Delta \Phi(f)$ for each detector, following the arguments of \citet{Wynn2} and \citet{Read_2023}. In general, the results show that there is a greater chance of detection for low mass stars systems, with BSk22 in particular being the most promising model. 

While our results may not be overly promising, they highlight the importance of stratification and composition when considering the g-mode spectrum of neutron stars. With the next generation of ground-based detectors having improved sensitivity at low frequencies, the future detection of g-modes remains a distinct possibility. With this in mind, the precise dependence on uncertain aspects of the underlying nuclear physics need to be further explored in future work.

\section*{Acknowledgements}

NA acknowledges generous support from STFC via grant number ST/Y00082X/1. PP acknowledges support from the Mar\'ia Zambrano Fellowship Programme (ZAMBRANO21), funded by the Spanish Ministry of Universities and the University of Alicante through the European Union's ``Next Generation EU'' package, as well as from the grant PID2021-127495NB-I00, funded by MCIN/AEI/10.13039/501100011033 and by the European Union, from the Astrophysics and High Energy Physics programme of the Generalitat Valenciana ASFAE/2022/026, funded by the Spanish Ministry of Science and Innovation (MCIN) and the European Union's ``Next Generation EU'' package (PRTR-C17.I1), and from the Prometeo 2023 excellence programme grant CIPROM/2022/13, funded by the Ministry of Education, Culture, Universities, and Occupation (Conselleria d'Educaci\'o, Cultura, Universitats i Ocupaci\'o) of the Generalitat Valenciana.

\section*{Data Availability}

Additional data related to this article will be shared on reasonable request to the corresponding author.

\bibliographystyle{mnras}
\bibliography{CowlingBibliography}

\end{document}